\documentclass[aps,prb,twocolumn,groupedaddress,floatfix]{revtex4-2} 
\usepackage{graphicx}
\usepackage{amsmath}
\usepackage{bm}
\usepackage{xcolor}

\usepackage{hyperref}
\begin{document} 

\title{Specific heat of Gd$^{3+}$ and Eu$^{2+}$-based magnetic compounds}

\author{D. J. Garc\'{\i}a}
\affiliation{CNEA-CONICET, GAIDI,Centro At\'{o}mico Bariloche, 8400 Bariloche, Argentina}

\author{J. G. Sereni}
\affiliation{CNEA-CONICET, GAIDI,Centro At\'{o}mico Bariloche, 8400 Bariloche, Argentina}

\author{A. A. Aligia}
\email{aaligia@gmail.com}
\affiliation{Instituto de Nanociencia y Nanotecnolog\'{\i}a, CNEA-CONICET, GAIDI, Centro At\'{o}mico Bariloche and Instituto Balseiro, 8400 Bariloche, Argentina}

\date{\today}

\begin{abstract}

We have studied theoretically the specific heat of a large number of non-frustrated magnetic structures described by the Heisenberg model for systems with total angular momentum $J=7/2$, corresponding to the 4f$^7$ configuration of Gd$^{+3}$ and Eu$^{+2}$.
For a given critical temperature (determined by the magnitude of the exchange interactions), we find that, to a high degree of accuracy, the specific heat is governed by two primary parameters: the effective number of neighbors $z$, which governs the extent of thermal and quantum fluctuations, and the axial anisotropy $K$.
The universality of $z$ (its ability to describe specific heat across diverse lattices) holds robustly for systems where exchange interactions do not strongly increase with distance and in the absence of frustration.
Otherwise, deviations from universality emerge. 
Using these two parameters we fit the specific heat of four Gd compounds and two Eu compounds, achieving a remarkable agreement. 
The present approach enables the extraction of magnetic interaction parameters not accessible through mean-field theory, offering a powerful tool for interpreting specific heat data in 4f$^7$ systems. 

\end{abstract}

\maketitle

\section{Introduction}
 \label{sec:intro}
Rare-earth-based materials are of great interest in both basic and applied condensed matter physics. These materials exhibit a wide range of fascinating physical phenomena. For example, they show  unconventional superconductivity\cite{Mathur98,Shioda21}, Kondo effect  \cite{Kobayashi08,Romero14,Magnavita16}, quantum criticality \cite{Alvarez04}, quadrupolar order and frustration\cite{Watanuki05,Okuyama05,Ji07,Song20,Franco24}.

 The specific heat of magnetic materials containing rare earths is typically highly sensitive to the environment surrounding the rare earth ions. This sensitivity arises from the splitting of the 4f orbitals by the crystal field and the consequent splitting of the total angular momentum $J$ due to strong spin-orbit coupling. Consequently, excitation energies are influenced by the surrounding environment. This, in turn, affects the temperature dependence of specific heat. 
  
 However, an exception is expected for compounds containing 
 Gd$^{+3}$ and Eu$^{+2}$. Both of these ions correspond to the 4f$^7$  configuration. 
 In this case, the ground-state multiplet, constructed by Hund's rules, is $^{8}S_{7/2}$. 
 This corresponds to a total spin  $S=7/2$, total orbital angular momentum $L=0$ (insensitive to crystal field), and total angular momentum $J=7/2$.

We show in this manuscript that the specific heat of 4f$^7$ systems can be characterized primarily by two parameters: the effective number of neighbors $z$, which accounts for the number of interaction bonds of each ion, and axial anisotropy $K$, which arises due to a small
admixture with $L=1$ states.

The thermodynamic behavior of magnetic systems is traditionally analyzed through lattice-specific models, but here we demonstrate that the effective number of neighbors $ z $ can provide a universal description of specific heat across diverse magnetic structures. This universality holds robustly for systems where exchange interactions do not strongly increase with distance and in the absence of frustration. 
In this parameter regime, the specific heat curves of different lattices with the same $ z $ collapse onto a single reference curve, enabling a simplified yet accurate interpretation of experimental data.  
Outside this regime, while the universality of $ z $ still provides a qualitative approximation,  deviations arise. 
These deviations highlight the importance of lattice geometry and frustration in systems where the hierarchy of exchange interactions is inverted.  
Our work builds on these insights to demonstrate that the two-parameter model (defined by $ z $ and axial anisotropy $ K $) successfully captures the specific heat of 4f$^7$ compounds (Gd$^{3+}$ and Eu$^{2+}$) across a wide range of magnetic structures.

Early measurements of
$C(T)$ on GdNi$_5$ \cite{GdNi5} were compared with calculations in the molecular field approximation for $T < T_C = 32$ K, where $T_C$ is the
critical temperature, with good agreement. However, the authors recognized
evident discrepancies at low temperature and in the vicinity of $T_C$. These discrepancies were attributed to ``short-range ordering effects not taken into account''. Simultaneous measurements of $C(T)$ on other Gd
compounds \cite{BouvierI} show more significant deviations from the
mean-field result.
The origin of such deviations was investigated by introducing
modulations in the amplitude of the magnetic moments
within the mean-field approximation \cite{BlancoII}.
Such a
model describes different possibilities of the observed temperature
dependence of $C(T)$ for $T<T_C$. However, some compounds clearly escape from those predictions.  This is especially true for observed magnetic fluctuations right above $ T_C $. 

Recent research on Eu$^{2+}$-based compounds shows similar
shortcomings of the mean-field approach  \cite{EuPtSi3, 2-2-1}.  The fact that Eu$^{2+}$ has a much larger atomic volume suggests it may participate in the formation of distinct crystalline structures. This indicates that the observed deviations are an intrinsic characteristic of the magnetic interactions. They are not dependent on crystalline symmetries or the sign of the interactions.  Therefore,
for a quantitative explanation of the observed $C(T)$, it is necessary
to include thermal and quantum fluctuations and go beyond mean field.
The effect of these fluctuations was calculated recently using
cluster mean-field theory \cite{heydarinasab2024paramagnon}.

In the present study we compute the magnetic specific heat for a wide variety of non‑frustrated (i.e. all exchange bonds can be satisfied simultaneously) lattices using the Heisenberg Hamiltonian with distance‑dependent exchange couplings $H_0$  and a single‑ion anisotropy term $H_A$:
\begin{equation} \label{ham}
    H = H_0 + H_A, 
    \end{equation}
 where
 \begin{equation}
H_0 = \sum_{i, \delta} J_\delta \mathbf{S}_i \cdot \mathbf{S}_{i+\delta} / 2 
    \end{equation}
and  
 \begin{equation}
      H_A = K \sum_i [3S_{iz}^2 - S(S+1)]. 
    \end{equation}
$\mathbf{S}_i, S_{iz}$ denote the spin (or total angular momentum)  $\vert S\vert = 7/2$ operators at site $i$ and its projection, $\delta$ labels the non-equivalent space vectors connecting different spins at short distances (nearest and possibly further neighbors), and $J_\delta$ is the corresponding exchange interaction.
Couplings $J_\delta$ are ferromagnetic  if negative or antiferromagnetic if positive.
We use $S$ instead of $J$ to denote the operators to avoid confusion with the exchange interactions.

The theoretically studied non-frustrated magnetic structures correspond to either ferromagnetic interactions or non-frustrated antiferromagnetic ones.
We find that the specific  heat of all these structures can be characterized
to a large degree of accuracy by two parameters: i) the effective number of neighbors (to be defined more precisely below) and ii) the magnetic anisotropy defined by $H_A$.  
Using these two parameters,  and the known value of $T_C$, we fit the data for the specific heat of GdNi$_3$Ga$_9$, GdPbBi, GdCu$_2$Ge$_2$, GdNiSi$_3$, Eu$_2$Pd$_3$Sn$_3$, and EuPdSn$_2$. 
The corresponding results are presented in Section \ref{expe}. 

The remainder of the paper is organized as follows. In Sec.\ref{sec:zeff} we define the effective number of neighbours
$z$ and discuss its physical meaning. Sec.\ref{sec:Anisotropy} derives the axial anisotropy term for 4f$^7$ ions. Section~\ref{method} describes the quantum Monte Carlo implementation and the finite‑size analysis. In Sec.\ref{nei} we present the specific‑heat curves for a large set of lattices and demonstrate the universality of $z$ and its limitations in the case of uniform couplings.
In Sec. \ref{aniso} we show the effect of anisotropy on the specific heat.
In Sec. \ref{compa} we analyze several lattices with non-uniform exchange couplings.
Section\ref{expe} contains the fits to six experimental compounds, and Sec.~\ref{sum} summarizes our findings.

In Appendix \ref{fluc} we discuss
the relation of $z$ with the quantum fluctuations.
Finally, quantitative measures of the differences between specific-heat curves are presented in  Appendices~\ref{app:distance}, \ref{app:distanceJsUniform}, \ref{app:distanceJsNonUniform}, and \ref{app:distanceExperiment}.

\section{The two main parameters}

\subsection{The effective number of neighbors}
\label{sec:zeff}
The effect of fluctuations on the specific heat of a spin system described by the Heisenberg model is expected to diminish as the number of neighboring spins increases.
The key insight is that fluctuations in the effective field experienced by each spin depend not just on the geometric coordination number, but on the relative strengths of different exchange pathways. 
For a given spin, the effective field is $\sum_{\delta} J_{\delta}\mathbf{S}_{i+\delta}$, where the sum runs over all neighbors. 
The relative fluctuation of this field should scale with the standard deviation divided by the mean value. 

We can draw an analogy to a binary random distribution (which would correspond to spin 1/2). It is known that after $N$ attempts in a binary random distribution $x$ with probabilities $p$ and $1-p$ for the values $x=1$ and $x=-1$, respectively, the mean value (proportional to the magnetic moment) is $\langle x \rangle = (2p-1)N$, and the standard deviation is $\sqrt{\langle x^2 \rangle - \langle x \rangle^2} = 2\sqrt{Np(1-p)}$. Thus, the ratio between the standard deviation and the mean value decreases as $1/\sqrt{N}$. Similarly, in our quantum Heisenberg model, one expects that as the number of neighbors increases, the specific heat is more similar to the mean-field result, in which thermal and quantum fluctuations are neglected and the effective magnetic field that each ion feels is replaced by its mean value.

Generalizing the above results to the case of a distribution with intensities $\pm |J_\delta|$ for $z_\delta$ neighbors, we obtain after some algebra (see Appendix \ref{fluc}):

\begin{equation}
\frac{\langle x^2 \rangle - \langle x \rangle^2}{\langle x \rangle^2} = \frac{4p(1-p)}{(2p-1)^2} \frac{1}{z}, \quad \mathrm{where} \quad \frac{1}{z} = \frac{\sum_{\delta} z_\delta J_\delta^2}{\left(\sum_{\delta} z_\delta |J_\delta|\right)^2} , \label{Rdefinicion}
\end{equation}

Here, $z_\delta$ represents the number of neighbors of type $\delta$ (with exchange interaction $J_\delta$), and the summation goes over all types of neighbors. The  quantity $z$  quantifies the effective number of neighbors, taking into account the strength ($|J_\delta|$) and number ($z_\delta$) of each type of interaction. 
Remarkably, for unfrustrated systems, the observed specific heat depends mainly on this effective number of neighbors $z$, largely independent of the specific lattice structure. 
$z$ quantifies the relative strength of spatial fluctuations.
For larger $z$ (higher coordination number), the results approach the mean-field ones, where thermal and quantum fluctuations are suppressed. Conversely, for lower $z$ (fewer neighbors),  fluctuations become more important, leading to a smaller specific heat below the critical temperature $T_C$ and a larger specific heat above $T_C$. 
Notably even for $z$ as large as $26$ there is a large departure of the quantum specific heat respect to the mean field result.

We note that $ z $ is distinct from the effective coordination number $ \hat{z} $ commonly used in inorganic and organic chemistry \cite{Mller1992InorganicSC}, where contributions from neighboring atoms are summed with weights decaying exponentially with distance. 
In contrast, our definition of $ z $ assigns weights determined by the exchange constants $ J_\delta $, which may be enhanced at longer distances and are not bound by geometric decay laws. 
Thus, $ z $ captures the dynamical influence of exchange interactions on spin fluctuations, rather than the static geometry of atomic arrangements.

\subsection{Axial anisotropy in 4f$^7$ ions}\label{sec:Anisotropy}
 Since neither the spin nor terms with $L=0$ are affected by crystal fields, one expects that the  only source of splitting of the ground-state octuplet is the  exchange interaction between rare-earth ions responsible  for the magnetic structure of the compound. 
 This holds as a first approximation. 
 In fact, we will show  that the specific heat of five Gd compounds can be well fitted ignoring crystal-field effects.
 
Nevertheless, due to spin-orbit coupling, the ground-state multiplet acquires a small admixture of the $^{6}P_{7/2}$ state with $L=1$. An explicit calculation for Gd$^{+3}$ in Appendix B of Ref. \onlinecite{Betancourth19} estimates that the excited multiplet contributes approximately $3\%$ to the ground state.
This admixture gives rise to an $L=1$ component in the ground state,
which enables its splitting in the presence of crystal fields.
This occurs if the symmetry at the rare-earth site is not high.   Specifically, if there is only a single axis of higher order, the anisotropy term $ H_A = C[3L_z^2 - L(L+1)] $ becomes relevant.

In high-symmetry environments such as tetragonal, cubic, or octahedral point groups ($T$, $T_d$, $T_h$, $O$, $O_h$), the $L=1$ states remain unsplit. This result, derived from group theory, is intuitively understood. A basis for $L=1$ states can be chosen to transform like the unit vectors
$\hat{x}$, $\hat{y}$, $\hat{z}$, which are equivalent in high-symmetric environments.
In contrast, if the maximum order axis of the point group is 2,
a term $L^2_{x}-L^2_{y}$ is also allowed.
In even lower symmetries, $L_{x} L_{y}$ may arise. However, their intensity is expected to be much smaller. Thus, experimental data can be fitted without including them. 

Using the Wigner-Eckart theorem, $H_A$ can be written in the form $H_A=K[3J^2_{z}-J(J+1)]$, where $K$ is of order $\sim 3\% C$.

\section{Methods}
\label{method} 

We have calculated the specific heat of a large number of magnetic structures.
The approach used a two-point numerical differentiation to compute the derivative of the internal energy as a function of temperature.
The internal energy was calculated using quantum Monte Carlo (QMC) simulations with the default ``minbounce'' algorithm (implemented in the ``dirloop\_sse'' application in the ALPS libraries \cite{albuquerque2007alps,bauer2011alps}). 
QMC allows us to handle large system sizes and complex lattices while properly accounting for thermal and quantum fluctuations that mean-field approaches miss.
Simulations were performed on systems of up to $8 \times 30^3=216000$ magnetic moments.
A typical simulation required approximately 10000 CPU hours per lattice on a standard workstation.
Classical Monte Carlo simulations fail to capture the low-temperature physics of quantum spin systems. By neglecting spin quantization and the Bose-Einstein statistics of magnons, they yield two hallmark errors: a finite residual entropy at $ T=0 $ (violating the third law) and a divergent $ C/T $ as $ T \to 0 $.
This artifact, documented in Ref.~\cite{kormann2010rescaled}, persists up to $ \sim 0.75\,T_C $ and obscures the magnetic signal in the temperature range $ 0.25 < T/T_C < 1.25 $ critical for comparison with experiment. 
Quantum Monte Carlo, by contrast, correctly enforces spin quantization and recovers $ C \to 0 $ as $ T \to 0 $, enabling a faithful representation of the full specific heat profile across all temperatures.

To generate the lattice configurations for the ALPS simulations, we employed the ``SUNNY'' package  \cite{dahlbom2024quantum} in conjunction with crystallographic information files (CIF) obtained from public repositories. 
SUNNY reads the CIF file and outputs the list of neighbour vectors up to a chosen cutoff.
Specifically, CIF files for Fe in simple cubic (SC), body-centered cubic (BCC), hexagonal close-packed (HCP), and face-centered cubic (FCC) structures, as well as C in diamond, MgAgAs (for GdPbBi),  Th(CrSi)$_2$ (for GdCu$_2$Ge$_2$), ErNiSi$_3$ (for GdNiSi$_3$), Mg$_2$MnGa$_3$ (for Eu$_2$Pd$_3$Sn) and MgAl$_2$Cu (for EuPdSn$_2$) were retrieved from the Materials Project \cite{10.1063/1.4812323} database. The ErNi$_3$Al$_9$ CIF file was obtained from the Crystallography Open Database \cite{Grazulis2009} using data from Ref.  \onlinecite{Gladyshevskii:du0339}.
Lattice parameters for GdCu$_2$Ge$_2$ were taken from Ref. \cite{rieger1969ternary}, for GdPbBi from Ref. \cite{haase2002equiatomic}, for GdNiSi$_3$ from Ref. \cite{Arantes18}, for Eu$_2$Pd$_3$Sn from Ref. \cite{akbar2023}, and for EuPdSn$_2$ from Ref. \cite{Curlik18}. In each case, the magnetic structure was constructed by assigning a spin moment $S = 7/2$ to the respective atoms (Gd or Eu).
For all lattices, except HCP, we used a rectangular cuboid supercell geometry as the conventional unit cell, resulting in varying numbers of sites within the supercell, with the diamond lattice requiring the largest (8 sites). Simulations were performed on systems containing $N^3$ supercells with periodic boundary conditions, typically with $N = 10$, $20$, and $30$. For the critical temperature, results were extrapolated to the thermodynamic limit assuming a $1/N$ dependence.

The well-known sign problem \cite{Pan_2024} associated with QMC simulations limited our investigation to non-frustrated magnetic structures. At low temperatures, the number of Monte Carlo steps required for statistically reliable results increases exponentially. Consequently, the internal energy was primarily computed in a temperature range between  $0.2T_C$ and $2T_C$, where $T_C$ is the critical temperature of the system. In a few cases,  simulations were extended to $0.1T_C$ at low temperatures. However, the reliability of the results  below $0.2T_C$ is limited due to finite-size effects.
We used $2 \times 10^5$ Monte Carlo steps for thermal equilibration and $4 \times 10^6$  steps for measurements.
These parameters results in a well converged energy and a smooth specific heat. 
The resulting error bars (not shown in figures) are close to the line width.

The coordination number (number of neighbors included) for each lattice is provided in parentheses after its name. For example, FCC(18) refers to a face-centered cubic lattice with 12 first neighbors and 6 second-nearest neighbors. 
Additional neighbor shells were included by increasing the cutoff radius $d$ for neighbor interactions.

Our theoretical results for the specific heat of all lattices used here are available in Ref. \onlinecite{note}.

\section{Role of the effective number of neighbors}
\label{nei} 

\begin{figure}[ht]
\begin{center}
\centering
\includegraphics[width=\columnwidth]{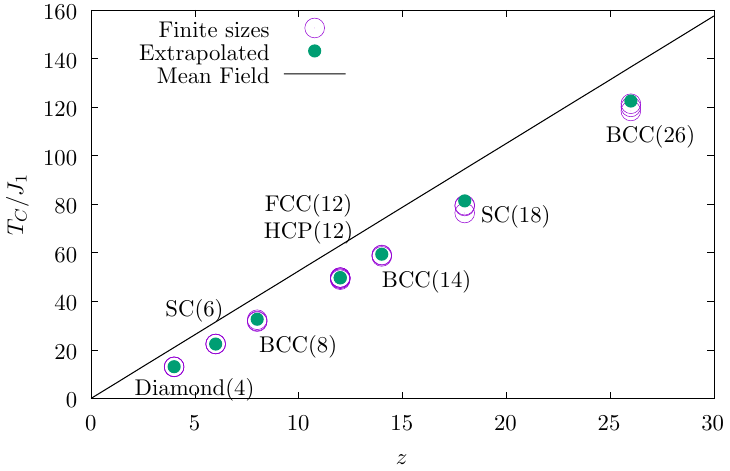} 
\caption{Critical temperature $T_C$ in units of the nearest‑neighbour exchange as a function of effective number of neighbors $z$ for several ferromagnetic structures. Finite-size results ($N=10,20,30$) and extrapolated values are shown, with the mean-field value $T_C^\mathrm{MF}$ indicated by the straight line.}
\label{tc}
\end{center}
\end{figure}

In this section we report our results for the specific heat and critical temperature for several simple lattices as a function of the number of neighbors that interact with a given site.
For simplicity, in all lattices considered, all sites are equivalent, and we assume all $J_\delta=J_1$, where $J_1$ is the  nearest-neighbor (NN)  interaction, up to a certain distance $d$ and 0 for longer distances.
Therefore, the coordination number given by Eq. (\ref{Rdefinicion}) reduces to $z=\Sigma _{\delta}z_{\delta}$.
$z$ is given by the total number of neighbors within a sphere of radius $d$. 
In Section \ref{compa} we discuss more general cases 
demonstrating that the results are not severely affected by this 
assumption.

Most of our studies assume ferromagnetic interactions 
($J_\delta <0$). Generally, the results for the corresponding non-frustrated antiferromagnetic structures are very similar, particularly for $z>6$,  except at very low 
temperatures, where the specific heat is dominated by magnons and scales as $T^{3/2}$ or  $T^{3}$ for ferro- and antiferromagnetic structures respectively (which can be seen in Fig. \ref{figure:ChangeJsGreaterJ1}, Sec. \ref{compa}).

In Fig. \ref{tc}, we show the critical temperature $T_C$ as a function of $z$ for several ferromagnetic structures, including finite-size results and an extrapolated value. Calculations have been done in finite systems containing 
$N^3$ conventional unit cells, with $N=10$, 20, and 30, as explained in Section \ref{method}.
The extrapolated value is also indicated in the figure.

As expected, $T_C$ increases almost linearly
with $z$ as the mean-field value $T_C^\mathrm{MF}$ (the effective 
magnetic field due to the interactions is proportional to $zJ_1$,
see Section \ref{MF}). But it is lower than $T_C^\mathrm{MF}$,
particularly for low $z$ due to the effect of thermal and quantum fluctuations, which is larger for low $z$ (a similar behavior was reported for $S=1/2$ in Ref. \cite{gonzalez2023finite}).

For large enough $ z $, one expects that $ T_C $ approaches $ T_C^{\mathrm{MF}} $.
However, in finite systems with periodic boundary conditions, the coordination number decreases when the interaction range $d$ becomes comparable to the system size, because neighbors at different distances overlap. 
This might prevent us from reaching the thermodynamic limit with our method.
Arguing as in the introduction that fluctuations decrease as  $1/z$ , one should expect that the critical temperature approaches the MF result much faster than the results shown on Fig. \ref{tc}.

\begin{figure}[ht]
\begin{center}
\centering
\includegraphics*[width=\columnwidth]{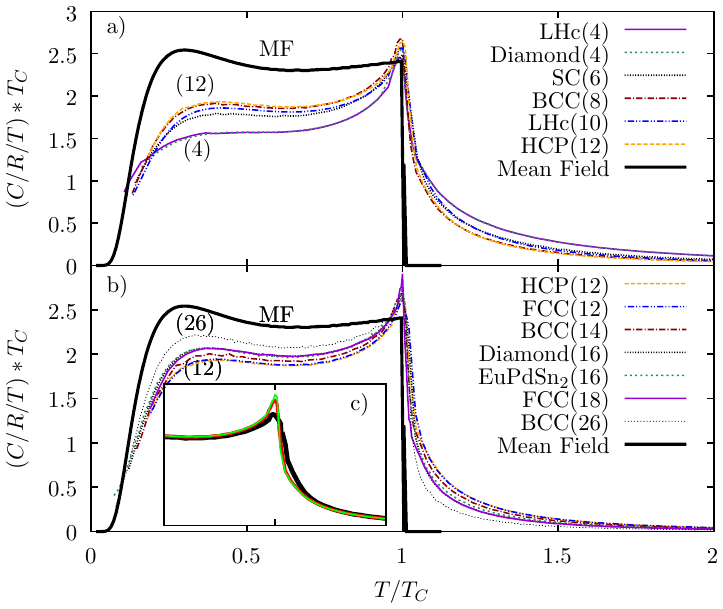}\\ 
\end{center}
\caption{Specific heat $C/T$ versus reduced temperature $T/T_C$ for several ferromagnetic structures with different effective number of neighbors $z$. Top panel ($z \leq 12$) and bottom panel ($z \geq 12$) show distinct ranges. The curves for SC, FCC, and HCP structures coincide at $z=18$, while SC, BCC, and EuPdSn$_2$ coincide at $z=26$, explaining the single curves shown for each set. LHc(4) denotes the layered honeycomb lattice with $z=4$. The bottom inset shows finite-size effects for HCP(12) with $N=10$ (black), $N=20$ (red), and $N=30$ (green) in the range $0.5 < T/T_C < 1.5$.
}
\label{cst}
\end{figure}

In Fig. \ref{cst} we show the results for 
the specific heat $C$ as a function of  temperature
for a large number of ferromagnetic structures with 
different numbers of magnetic neighbors $z$ (the number of nearest neighbors up to third order is shown in Table \ref{Tabla:Vecinos} and its critical temperatures are shown in Table \ref{Table:CriticalTemperatures}).
Also shown is the mean-field result which corresponds to the limit $z \rightarrow \infty$.
This result shows larger $C$ for $T<T_C$, while $C=0$ for $T>T_C$. 
In contrast,  the specific heat for a honeycomb layered system with $z=4$ (three nearest neighbors within the plane of the honeycomb layer and one above or below the layer in alternating sites, LHc(4)) shows the smallest specific heat below $T_C$ and the largest one above $T_C$. 
This behavior arises from short-range correlations, which are absent in mean-field theory and persist for $T > T_C$, thereby contributing to the specific heat. 
The importance of these correlations increases as $z$ decreases. 
Since the total entropy per site is fixed at $\ln(7)$ (for $S=7/2$), this redistribution of entropy implies that $C/T$ decreases with decreasing $z$ for $T < T_C$. 
High-temperature series expansions \cite{gonzalez2023finite} confirm that, well above $T_C$, the specific heat of three-dimensional Heisenberg magnets should exhibit a $1/T^3$ tail, consistent with our observed behavior.
Note that the diamond structure, which is distinct from the previous one but also has $z=4$, has a very similar specific heat.
The low-temperature shoulder is a Schottky-like effect due to the internal field \cite{Facio15,BlancoII}.
The lower panel inset shows that finite size effects are only appreciable near $T_C$ as exemplified in the case of the HCP(12) lattice. 
For this reason, we show specific heat curves from QMC corresponding to $N = 20$.  

Increasing $z$ from 4 to 6, the specific heat of the simple cubic structure with six NN displays the expected trend of increasing (decreasing) $C$ below (above) $T_C$.
Increasing $z$ up to 14 using the BCC structure with 6 NN and 8 next-nearest neighbors (NNN), the same trend is observed in general. 
However, the specific heat of the BCC lattice with $z = 8$ (first neighbors) lies very close to the cubic FCC with $z = 12$  and above $z=10$. 
The FCC lattice with $z=12$ has a $C$ quite close to the HCP with $z=12$ (first and second neighbors).

In Fig.  ~\ref{figure:HeatMapJsIguales} of Appendix \ref{app:distanceJsUniform} we quantify the differences between the specific heat curves using the distance measure defined in Appendix \ref{app:distance}.

The diamond(16) and EuPdSn$_2$(16) lattices when 16 neighbors are considered also have very similar specific heat.
Other exception correspond to the FCC(18) case composed  of  twelve NN and  six NNN that has $C$ quite close to the previous mentioned cases of  diamond(16) and EuPdSn$_2$(16).
Nevertheless,  the specific heat is very similar for different structures with the same $z$. 
Even if the mean-field theory predicts identical behavior  for all  coordination numbers, without accounting for lattice geometry, this universality seen with QMC across different lattice types with the same $z$ is remarkable.

\begin{table}
\centering
 \begin{tabular}{|c|c|c|c|}
\hline
Lattice & 1st NN ($z_1$) & 2nd NN ($z_2$) & 3rd NN ($z_3$)  \\ \hline
EuPdSn$_2$      & 2 & 2 & 4 \\
\hline
Eu$_2$Pd$_3$Sn$_3$      & 1 & 2 & 1 \\
\hline
GdNiSi$_3$      & 2 & 2 & 2 \\
\hline
GdCu$_2$Ge$_2$      & 4 & 4 & 8 \\
\hline
LHc      & 3 & 1 & 6 \\
\hline
Diamond      & 4 & 12 & 12 \\
\hline
SC      & 6 & 12 & 8 \\
\hline
BCC      & 8 & 6 & 12 \\
\hline
FCC      & 12 & 6 & 24 \\
\hline
HCP      & 12 & 6 & 2 \\
\hline
\end{tabular}

\caption{Number of neighbors at increasing distances for various lattice types.\label{Tabla:Vecinos}}
\end{table}

\begin{table}
\centering

\begin{tabular}{|c|c|c|c|c|c|c|c|}
\hline
Lattice & $ z_1 $ & $ z_2 $ & $ z_3 $ & $ J_1 $ & $ J_2 $ & $ J_3 $ & $ T_C $ (K) \\
\hline
LHc(4,AF)$^\dagger$ & 3 & 1 & 6 & 1 & 1 & 0 & 13.2$^\ddagger$ \\
\hline
LHc(4)$^\dagger$ & 3 & 1 & 6 & -1 & -1 & 0 & 13.1 \\
\hline
Diamond(4) & 4 & 12 & 12 & -1 & 0 & 0 & 13.1 \\
\hline
SC(6,AF)$^\dagger$ & 6 & 12 & 8 & 1 & 0 & 0 & 22.4 \\
\hline
SC(6) & 6 & 12 & 8 & -1 & 0 & 0 & 22.4 \\
\hline
BCC(8,AF)$^\dagger$ & 8 & 6 & 12 & 1 & 0 & 0 & 32.3$^\ddagger$ \\
\hline
BCC(8) & 8 & 6 & 12 & -1 & 0 & 0 & 32.6 \\
\hline
LHc(10) & 3 & 1 & 6 & -1 & -1 & -1 & 39.2 \\
\hline
HCP(12) & 12 & 6 & 2 & -1 & -1 & 0 & 49.6 \\
\hline
FCC(12) & 12 & 6 & 24 & -1 & 0 & 0 & 49.8 \\
\hline
BCC(14) & 8 & 6 & 12 & -1 & -1 & 0 & 59.4 \\
\hline
EuPdSn$_2$(16)$^*$ & 2 & 2 & 4 & -1 & -1 & -1 & 69.1 \\
\hline
Diamond(16)$^\dagger$  & 4 & 12 & 12 & -1 & -1 & 0 & 70.1 \\
\hline
FCC(18)$^\dagger$ & 12 & 6 & 24 & -1 & -1 & 0 & 79.6 \\
\hline
SC(18) & 6 & 12 & 8 & -1 & -1 & 0 & 79.8 \\
\hline
BCC(26) & 8 & 6 & 12 & -1 & -1 & -1 & 122. \\
\hline
\end{tabular}

\caption{Extrapolated critical temperatures computed with QMC for Hamiltonians in the lattices considered in the text.
Lattices marked with $*$ have more than third nearest-neighbors (see text). 
For lattices marked with $^\dagger$ we report the $N=20$ critical temperature. AF denotes that antiferromagnetic interactions were used.
Uncertainty in $T_C$ is below 2\%.
$^\ddagger$ correspond to a N\'eel temperature.\label{Table:CriticalTemperatures}}
\end{table} 

\section{Effect of anisotropy}
\label{aniso} 

In this section, we address the role of anisotropy first in the mean-field approximation and then incorporate the effect of  fluctuations.

\subsection{Mean-field Approximation}
\label{MF}

In this subsection, we discuss the effect of the term  $H_A$ on the specific heat calculated in the mean-field approximation.
In this approximation it is assumed that each site is subjected to a constant effective magnetic field
resulting from the interaction with the remaining sites.
Specifically, the interactions appearing in $H_0$ [see Eq. (\ref{ham})]
are approximated as
\begin{equation}
\mathbf{S}_{i}\cdot \mathbf{S}_{j}
\approx  \langle \mathbf{S}_{i} \rangle \cdot \mathbf{S}_{j}
+ \mathbf{S}_{i}\cdot \langle  \mathbf{S}_{j} \rangle
- \langle \mathbf{S}_{i} \rangle \cdot \langle \mathbf{S}_{j} \rangle,
\label{mfd}
\end{equation}

The mean-field approximation assumes  that all sites are equivalent (long-range magnetic order), which may not hold near the ordering temperature.
We assume that there is no external magnetic field. 
For simplicity, we restrict the analysis to the easy axis (uniaxial anisotropy or anisotropy along a single axis) case $K<0$, for which the spins point in the direction of the anisotropy axis $z$. 
In the next subsection, both signs of $K$ and the effects of thermal and quantum fluctuations will be considered.

For an antiferromagnetic order, we rotate 180 degrees the spins
pointing in the $-z$ direction around a perpendicular axis so that for all sites
$\langle S_{z} \rangle \geq 0$. This expectation value serves as the order parameter for the magnetic phase.

The effective magnetic field $B_{\mathrm{eff}}$ is given by
\begin{eqnarray}
Z&=\sum\limits_{M=-J}^{J}e^{-\beta E_{M}} \mathrm{,~}
\langle S_{z}\rangle& =\sum\limits_{M=-J}^{J}\frac{Me^{-\beta E_{M}}}{Z} \mathrm{,~} \nonumber \\
\tilde{B}&=g\mu _{B}B_{\mathrm{eff}}=I\langle S_{z}\rangle ,  \label{beff}
\end{eqnarray}
where $g$ is the gyromagnetic factor,  $\mu _{B}$ is the Bohr magneton, 
$I=|\Sigma _{\delta}z_{\delta}J_{\delta}|$ is the sum of all interactions ($z_{\delta}$ is the
number of neighbors at distance $\delta$ and 
$J_{\delta}$ its intensity), $\beta
=1/(k_{B}T)$ where $T$ is the temperature and $k_{B}$ is the Boltzmann
constant. The energies $E_{M}$ for each eigenvalue $M$ of  the operator
$S_{z}$ are given by:

\begin{equation}
E_{M}=-\tilde{B}M + K[3M^2-S(S+1)].  \label{em}
\end{equation}
Eqs. (\ref{beff}) and (\ref{em}) permit to determine 
$\tilde{B}$
self-consistently. In the limit $\tilde{B}\rightarrow 0$ one obtains an
equation for the critical temperature

\begin{equation}
k_{B}T_C^\mathrm{MF}=I\langle S^2_z \rangle \mathrm{, where~~}
\langle S^2_z \rangle= \sum \limits_{M=-J}^{J}\frac{M^{2}e^{-\beta E_{M}}}{Z}.
\label{tcf}
\end{equation}
The specific heat is obtained from the numerical derivative of the entropy $S$ with respect to temperature: 

\begin{equation}
C=T\frac{dS}{dT} \mathrm{,~~~ }S=\sum\limits_{M=-J}^{J}p_{M}\ln p_{M} \mathrm{,~~~ } p_{M}=\frac{e^{-\beta E_{M}}}{Z}  \label{c}
\end{equation}

\begin{figure}[ht]
\begin{center}
\centering
\includegraphics[width=\columnwidth]{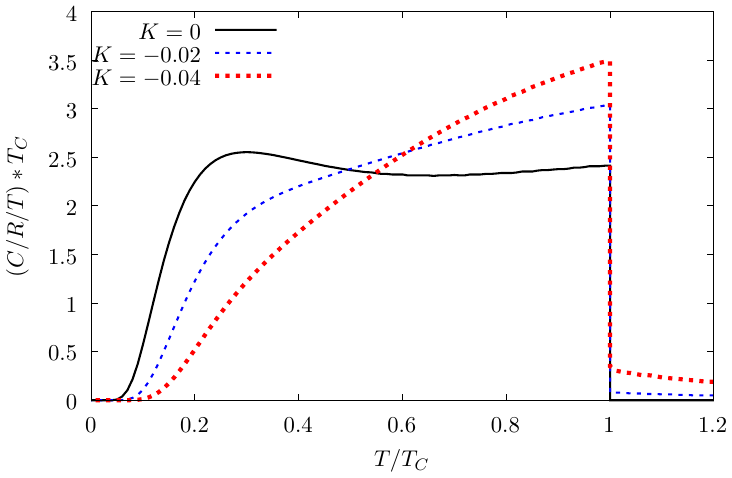}
\caption{Specific heat $C/T$ versus $T/T_C$ for the mean-field approximation with varying axial anisotropy $K$ (expressed in units of $k_B T_C$).}
\label{cstmf}
\end{center}
\end{figure}

In Fig. \ref{cstmf} we show the resulting effect of the uniaxial anisotropy
on the specific heat keeping the value of $k_B T_C=1$ constant, as the unit
of energy. For $K=0$, the value of the intensity for a given $T_C$ is
$I=3 k_B T_C/[S(S+1)]$, leading to $I=4/21 \sim 0.190$ for $k_B T_C=1$.

For $K=-0.02$, we obtain $I=0.1531$ from Eq. (\ref{tcf}). This is because the
expectation value $\langle S^2_z \rangle$ that enters Eq. (\ref{tcf}) is
larger for $K <0$ and should be compensated with a smaller $I$.
In addition, since the values of high $S_z$ projection $|M|$ are favored,
the effective magnetic field is more efficient than for $K=0$ in splitting
the lowest energy levels and therefore the entropy and specific heat decrease for $T < 0.4 T_C$  with respect to the isotropic
case. This is compensated by an increase in $C/T$ for $0.5 T_C < T <T_C$.

For $K=-0.04$, $I$ decreases further to $I=0.1263$, and the changes in the specific heat are more marked in the same direction. For large negative $K$, $I=1/(7/2)^2 \sim 0.0816$.

In the mean-field approximation, anisotropy shifts entropy from low to intermediate temperatures, with $T_C$ determined by the balance between exchange interactions and anisotropy.
Above $T_C$ and  when  $K\ne0$ the specific heat is different from zero reflecting the onsite energy fluctuations induced by the CEF term.

\subsection{Beyond Mean-field}
\label{bey}

\begin{figure}[ht]
\begin{center}
\includegraphics*[width=\columnwidth]{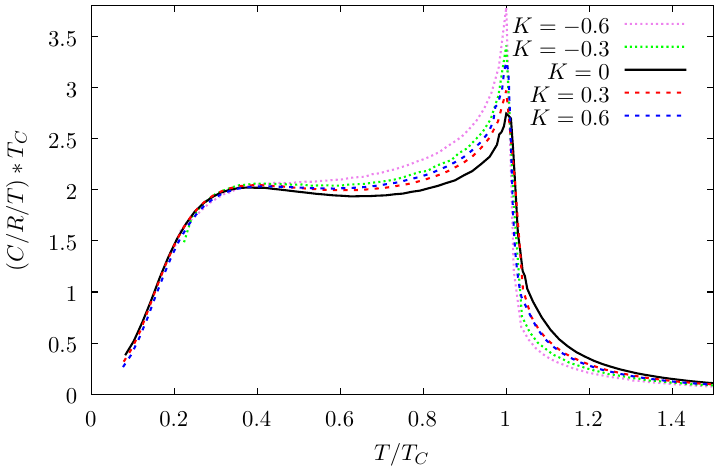}
\caption{ Specific heat $C/T$ versus $T/T_C$ for the EuPdSn$_2$(16) structure with varying axial anisotropy $K$ ($J_\delta = -1$). The critical temperatures for $K = -0.6, -0.3, 0, 0.3, 0.6$ are approximately $78, 74, 68.5, 71.5, 73$ respectively.
To compare with Fig. \ref{cstmf} the ratio $K/{k_B T_c}$ for these cases are $-.0077$, $-0.0040$,$0$,$0.0042$ and $0.0082$ respectively.}
\label{csta}
\end{center}
\end{figure}

In this subsection, we discuss the effect of anisotropy in cases where the fluctuations of the effective magnetic field  are also included. 
Fig. \ref{csta} shows the specific heat for the EuPdSn$_2$(16) structure (described in section \ref{expe}) with varying anisotropy $K$ for moderate values ($\vert K\vert \lesssim J_{\delta}$).
For $K = 0$, the system is isotropic. Negative $K$ values introduce axial anisotropy, favoring alignment along the $ z $-axis, whereas positive $K$ values
produce planar anisotropy, favoring alignment within the $xy$-plane.
Qualitatively, the results for $ T < T_C $ mirror those of the mean-field approximation, with anisotropy shifting entropy from the region $ T < 0.5 T_C $ to $ 0.5 T_C < T < T_C $, and the effect is more pronounced for negative $K$.

An anisotropy-induced energy gap arises from the splitting of spin states due to the axial anisotropy term $ H_A = K[3S_{z}^2 - S(S+1)] $. 
For negative $ K $, this term favors alignment along the $ z $-axis, creating a gap between low-energy $ |S_z| $ states with large $\vert S_z\vert$ values. 
This restricts spin fluctuations above $ T_C $, reducing entropy and specific heat. 
For positive $ K $, the anisotropy favors in-plane alignment, leading to a inverse distribution of low-energy states but with similar effects on the specific heat.
The QMC simulations explicitly capture these quantum and thermal fluctuations, which mean-field theory cannot account for. 

The Schottky-like shoulder observed in the specific heat is attributed to the internal magnetic field which creates a low-energy gap in the spin excitation spectrum.
This feature is particularly sensitive to the axial anisotropy. 
For positive $ K $, the low-energy states are more closely spaced, leading to a smaller effect compared to negative $ K $ where  low-energy states are more separated.
In this last case the Schottky-like shoulder is more affected by  the anisotropy.

The entropy removed above $T_C$ by the anisotropy term is recovered between the  Schottky-like shoulder and $T_C$.

\section{Comparison of Different Structures with the Same Effective Number of Neighbors}  
\label{compa}  
 
\begin{figure}[htbp]  
\begin{center}  
\includegraphics*[width=\columnwidth]{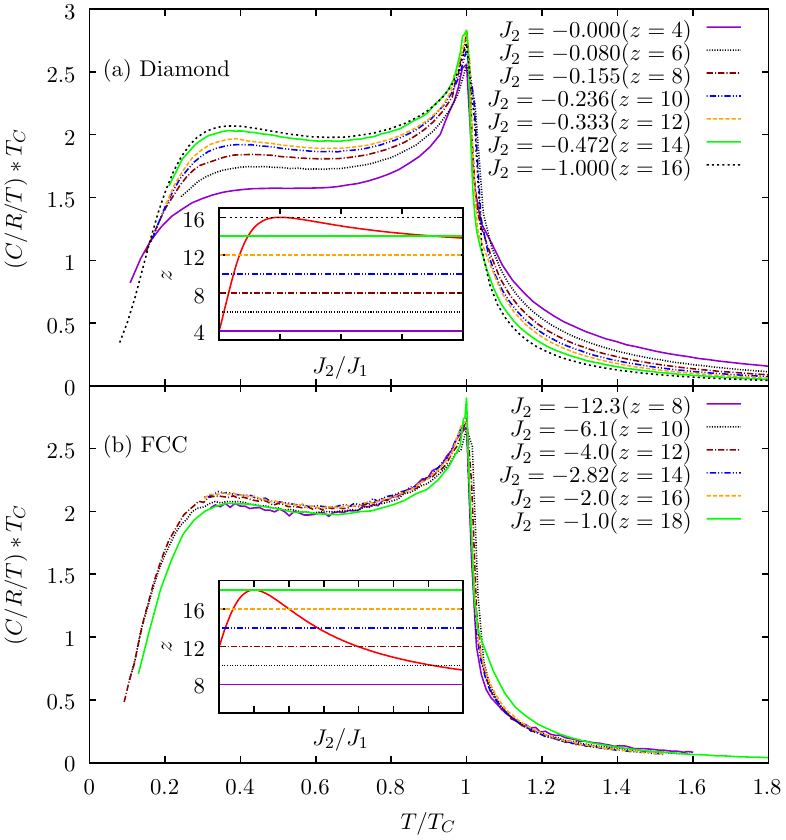}  
\end{center}  
\caption{Specific heat $C/T$ versus $T/T_C$ for  diamond and FCC lattices with different values of effective number of neighbors $z$ [defined by Eq. (\ref{Rdefinicion})], with $J_1 = -1$. The critical temperatures $T_C$ and $z$ values are tabulated in Table \ref{Table:CambioJs}. Top panel: Diamond lattice with varying interactions with $|J_2| \leq |J_1|$ ($z = 4, 6, 8, 10, 12, 14, 16$ from bottom to top). Top panel inset: $z$ as a function of $J_2$ for diamond. Bottom panel: same results for FCC lattices with varying interactions with $|J_2| \geq |J_1|$  ($z = 8, 10, 12, 14, 16, 18$). Bottom panel inset: $z$ as a function of $J_2$ for FCC.
}  
\label{figure:ChangeJs}  
\end{figure}  

Our central finding is that the specific heat of magnetic systems is primarily determined by the effective number of neighbors $z$,
defined by Eq. (\ref{Rdefinicion})
rather than the specific lattice structure or interaction pattern.
The parameter $z$ quantifies the relative strength of thermal and quantum fluctuations. 
Physically, $z$ represents the effective number of neighbors that meaningfully contribute to the exchange field experienced by each spin, accounting for both geometric coordination and relative interaction strengths.
In Section \ref{nei}, we analyzed the impact of varying $ z $ on the specific heat, while keeping $ J_\delta = J_1 $. 
Here, we extend this analysis by investigating how changes in $ J_\delta $ affect the specific heat, while maintaining the effective number of neighbors $ z $ constant [see Eq. (\ref{Rdefinicion})]. 

To illustrate this principle, we use diamond and FCC lattices. These two systems allow us to explore cases where $ z $ varies between 4 (just nearest neighbors) and 16 (nearest and next-nearest neighbors) for the diamond lattice, and between 8 and 18 for the FCC lattice.  
In Fig. \ref{figure:ChangeJs}(a) and (b), we show how the specific heat changes with $ J_2 $, for both lattices and for $z$ values ranging from 4 to 18. The insets in each panel display $z$ as a function of $J_2$ for the respective lattice.
For the diamond lattice with $ 0 \geq J_2 \geq J_1 $ ($ |J_2| < |J_1| $),
the specific heat curves closely follow the results shown in Fig. \ref{cst}.
For the FCC lattice with
$ |J_2| > |J_1| $, the specific heat appears almost independent of $ J_2 $. This behavior may be attributed to the FCC lattice’s non-bipartite nature, which could involve higher-order effective interactions that equilibrate fluctuations from first and second neighbors.

\begin{figure}[htbp]   
\begin{center}  
\includegraphics*[width=\columnwidth]{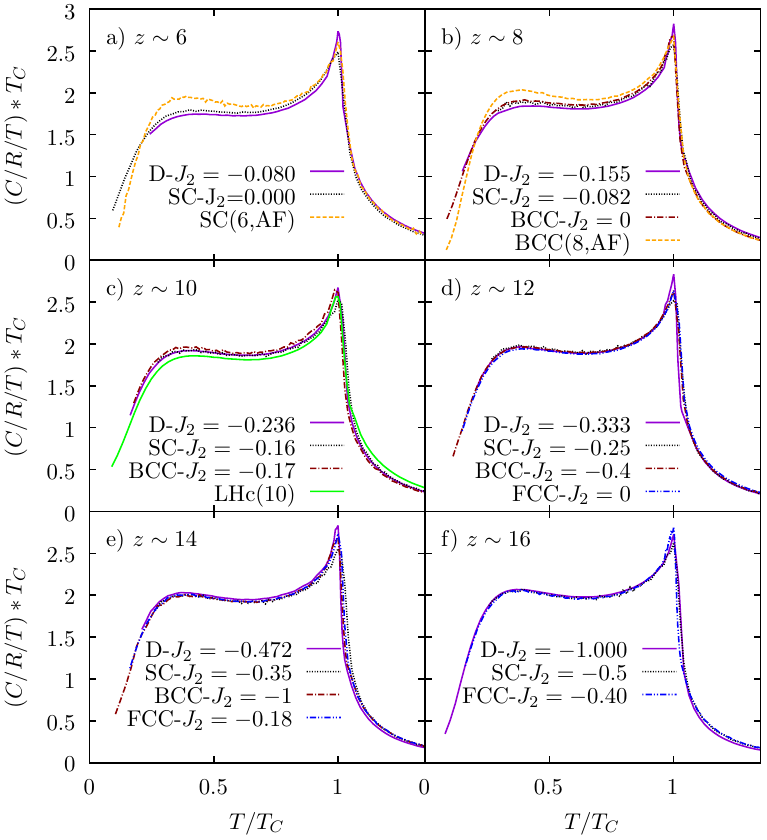}  
\end{center}  
\caption{Specific heat $C/T$ versus $T/T_C$ for various lattice structures with different values of the effective number of neighbors $z$ [defined by Eq. (\ref{Rdefinicion})], with $J_1 = -1$ and $0\geq J_2 \geq J_1$. The critical temperatures $T_C$ and $z$ values are tabulated in Table \ref{Table:CambioJs}. Purple lines correspond to diamond lattice results while black, brown,  blue and green correspond to simple cubic, BCC, FCC and LHc respectively. Panels a) to f) correspond to $z$ from 8 to 16. Orange lines on panels a) and b) correspond to antiferromagnetic interactions.
}  
\label{figure:ChangeJsGreaterJ1}  
\end{figure}  

A detailed comparison between diamond, simple cubic (SC), body-centered cubic (BCC), FCC, and layered honeycomb (LHc) lattices is shown in Figure \ref{figure:ChangeJsGreaterJ1}. In this figure, we present results for $ z $ values between 6 and 16 with $ |J_2| \leq |J_1| $. The agreement of the specific heat for different lattices with varying $ J_2 $ but the same $ z $ is excellent.
We also show with orange lines the specific heat for non-frustrated antiferromagnetic cases with $ z = 6 $ and $ z = 8 $, showing a clear convergence toward the ferromagnetic results as $ z $ increases.  

\begin{figure}[htbp]   
\begin{center}  
\includegraphics*[width=\columnwidth]{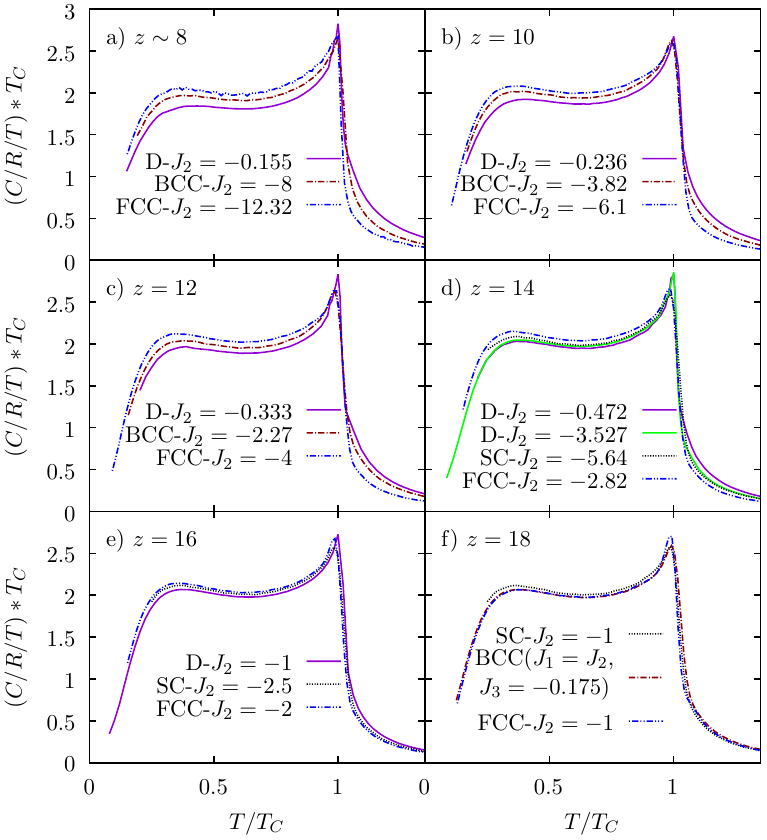}  
\end{center}  
\caption{Specific heat $C/T$ versus $T/T_C$ for various lattice structures with different values of effective number of neighbors $z$ [defined by Eq. (\ref{Rdefinicion})], with $J_1 = -1$. The critical temperatures $T_C$ and $z$ values are tabulated in Table \ref{Table:CambioJs}. Purple lines correspond to diamond lattice results while black, brown,  blue and green correspond to simple cubic, BCC, FCC and LHc respectively. Panels a) to f) correspond to $z$ from 8 to 16. Orange lines on panels a) and b) correspond to antiferromagnetic interactions.
}  
\label{figure:ChangeJsLowerJ1}  
\end{figure}  

In Figure \ref{figure:ChangeJsLowerJ1}, we show the specific heat evolution when
$ |J_2| \geq |J_1| $. As a reference, the purple lines depict the specific heat for the diamond lattice with the corresponding $ z $ value for $ |J_2| \leq |J_1| $. We compare the diamond (green), BCC (brown), simple cubic (SC, black), and FCC (blue) lattices. For values of $ |J_2| $ much larger than $ |J_1| $ (panels a–c), there is a significant departure from the diamond reference. 
The fact that the FCC lattice (non-bipartite) shows the largest deviation compared to BCC and SC suggests that higher-order interactions could equilibrate fluctuations from first and second neighbors. In panel (f), we also show the BCC case with $ z = 18 $ and three couplings, where the agreement with the SC and FCC lattices is remarkable.  

Quantitative measures of the difference in specific heats are provided in Tables~\ref{tab:J2_less_J1} and \ref{tab:J2_greater_J1}  of Appendix
\ref{app:distanceJsNonUniform}

These results demonstrate that the effective number of neighbors $ z $ encapsulates the essential physics of thermal and quantum fluctuations for $ 0 \geq J_2 \geq J_1 $
($ |J_2| \leq  |J_1| $), and it provides a good approximation for $ |J_2| >  |J_1| $. This enables a universal description of specific heat across diverse magnetic structures, independent of their lattice geometry or interaction patterns.

\subsection{Getting information beyond mean field}

The key insight of our work is that the effective number of neighbors $ z $ [defined in Eq. (\ref{Rdefinicion})] is the dominant parameter determining the specific heat, rather than the specific lattice structure or interaction pattern, when $ 0 \geq J_2 \geq J_1 $. This allows us to extract valuable information about the underlying exchange interactions that goes beyond what can be obtained from mean-field theory alone.  

To illustrate this, consider the diamond lattice with varying $ J_2 $ (Fig. \ref{figure:ChangeJsGreaterJ1}). For different values of $ J_2 $, we obtain different effective numbers of neighbors $ z $. Crucially, when we compare diamond lattices with different $ J_2 $ values but the same $ z $ to reference lattices with uniform couplings (SC, BCC, FCC, LHc), we find that the specific heat curves match almost perfectly. This demonstrates that the specific heat is insensitive to the detailed interaction pattern as long as $ z $ is the same.  

This observation provides a powerful tool for experimentalists. By comparing experimental specific heat data to our reference curves (available in Ref. \onlinecite{note}), one can extract the effective number of neighbors $z$ and, from $z$, infer the relative strengths of different exchange interactions. This is a significant advance over mean-field analysis, which cannot distinguish between different interaction patterns that yield the same critical temperature.

Let's examine how this works in practice using the examples from Fig. \ref{figure:ChangeJsGreaterJ1}. Assume that an experimental specific heat curve coincides with one of those displayed in the figure, while the exchange interactions $J_\delta$ are initially unknown.

\begin{enumerate}
 \item  Diamond lattice with $ J_2 = -0.080 $ ($ z = 6.03 $): 
   The specific heat matches SC(6). Using Eq. (\ref{Rdefinicion}) with $ z_1 = 4 $ and $ z_2 = 12 $ for the diamond lattice, we solve for $| J_2/J_1 |$:  
 \[
 1/z = \frac{z_1 J_1^2 + z_2 J_2^2}{(z_1 |J_1| + z_2 |J_2|)^2} = 1/6.
 \]
 This yields $| J_2/J_1| \approx 0.0787 $, very close to the actual value of $ 0.080 $. This demonstrates that the parameter $z$ can be used to extract the ratio of interaction strengths with high accuracy. Using the critical temperature as an energy scale, we can even determine the absolute values of $ J_1 $ and $ J_2 $.

\item Diamond lattices with $J_2 = -0.155$ ($z = 8$) and $J_2 = -0.333$ ($z = 12$): Both match BCC(8) and FCC(12) as specific heat for these reference lattices are very close. 
This indicates that these diamond lattices require next-nearest-neighbor interactions for an accurate description. The analysis limits the possible parameter sets to $|J_2/J_1| \approx 0.155$ or $|J_2/J_1| \approx 0.333$, providing valuable constraints on the interaction strengths.

 \item BCC lattice with $ J_2 = -0.40 $: As in the previous case BCC($J_2 = -0.40$)  matches BCC(8) and FCC(12) and lead to multiple possibilities: if $z = 8$, then $|J_2/J_1| = 0$ or $8$; if $z = 12$, then $|J_2/J_1| = 0.390$ or $2.276$.  This illustrates how
$ z $ can narrow down plausible interaction ratios, even when multiple solutions exist.

 \item BCC lattice with $ J_1 = J_2 = -1.0 $, $ J_3 = -0.175 $: The specific heat matches FCC(18). This indicates that at least both nearest-neighbor (8) and next-nearest-neighbor (6) interactions are relevant. Considering interactions up to third order leads to a continuum of possibilities  that can be further constrained by additional data.
\end{enumerate}

The remarkable universality of specific heat curves across different lattice types with the same $z$ suggests that the thermodynamics of these systems is primarily governed by the statistical properties of the exchange field distribution rather than detailed lattice geometry.
As long as the exchange interactions are unfrustrated  and the effective field distribution has the same variance-to-mean ratio (characterized by $z$), the thermodynamic behavior becomes universal. 

This approach provides a systematic way to extract information about the underlying exchange interactions from specific heat measurements. It goes beyond mean-field theory by providing  bounds for the number of relevant couplings and, in some cases, a limited set of options for interaction strengths. This is particularly valuable for experimentalists who often have limited information about the detailed magnetic interactions in their materials.

\begin{table}
\centering
 \begin{tabular}{|c|c|c|c|c|c|}
 \hline Lattice & $ J_1 $ & $ J_2 $ & $ J_3 $ & $z$ & $ T_C $ (K) \\ \hline
 Diamond & -1.0 & -0.080 & 0.0 &6.03 & 18.1 \\ \hline 
 Diamond & -1.0 & -0.155 & 0.0&8 & 22.8 \\ \hline
 Diamond & -1.0 & -0.236 & 0.0&10 & 27.6 \\ \hline
 Diamond & -1.0 & -0.333 & 0.0&12 & 33.4 \\ \hline
 Diamond & -1.0 & -0.472 & 0.0&14 & 41.4 \\ \hline
 Diamond & -1.0 & -3.527 & 0.0&14 & 202. \\ \hline
 SC & -1.0 & 0.0 & 0.0 & 6 & 22.5 \\  \hline 
SC & -1.0 & -0.082 & 0.0 & 8 & 27.3 \\  \hline 
SC & -1.0 & -0.16 & 0.0 & 10 & 31.9 \\  \hline 
SC & -1.0 & -0.25 & 0.0 & 12 & 37.3 \\  \hline 
SC & -1.0 & -0.35 & 0.0 & 14 & 42.8 \\  \hline 
SC & -1.0 & -0.5 & 0.0 & 16 & 51.8 \\  \hline 
SC & -1.0 & -2.5 & 0.0 & 16 & 159.6 \\  \hline 
SC & -1.0 & -5.64 & 0.0 & 14 & 322.4 \\  \hline 
BCC & -1.0 & -0.17 & 0.0 & 10 & 37.0 \\ \hline 
BCC & -1.0 & -0.40 & 0.0 & 12 & 43.1 \\ \hline 
BCC & -1.0 & -2.27 & 0.0 & 12 & 92.2 \\ \hline 
BCC & -1.0 & -3.82 & 0.0 & 10 & 130.1 \\ \hline 
BCC & -1.0 & -8.0 & 0.0 & 8 & 230. \\ \hline 
BCC & -1.0 & -1.0 & -0.175 & 18 & 70.8 \\ \hline 
FCC & -1.0 & -0.18 & 0.0 & 14 & 55.0 \\ \hline 
FCC & -1.0 & -0.40 & 0.0 & 16 & 62.2 \\ \hline 
FCC & -1.0 & -2.0 & 0.0 & 16 & 107.7 \\ \hline 
FCC & -1.0 & -2.82 & 0.0 & 14 & 129.7 \\ \hline 
FCC & -1.0 & -4.0 & 0.0 & 12 & 159.6 \\ \hline 
FCC & -1.0 & -6.1 & 0.0 & 10 & 211.4 \\ \hline 
FCC & -1.0 & -12.32 & 0.0 & 8 & 366.3 \\ \hline
\end{tabular}
\caption{Critical temperatures for Hamiltonian models with unequal interactions. All data corresponds to $N=20$.\label{Table:CambioJs}}
\end{table} 

\bigskip 

The practical significance of this finding is substantial. By comparing experimental specific heat to our reference curves, one can extract the effective number of neighbors $z$ and, from $z$, infer the relative strengths of different exchange interactions. 
This information is inaccessible through mean-field analysis alone and can guide the development of more accurate theoretical models. 
For example, if  the specific heat of a SC compound  matches our reference curve for $z = 8$ (thus it also matches $z=12$), we can immediately infer that the system likely has a combination of nearest-neighbor and next-nearest-neighbor interactions with a ratio $|J_2/J_1| \approx 0.08$ ($z=8$) or $|J_2/J_1| \approx 0.25$ ($z=12$).

For quantitative comparisons between specific heat curves across lattices, including distances under uniform and non-uniform couplings, as well as fits to experimental data, see Appendices~\ref{app:distance}, \ref{app:distanceJsUniform}, \ref{app:distanceJsNonUniform}, and \ref{app:distanceExperiment}.

This framework represents a significant advance in our ability to interpret magnetic specific heat data for 4f$^7$ systems, as demonstrated in Section \ref{expe} where we apply these principles to fit experimental data for six different compounds.

\section{Fits of experimental data}
\label{expe}

\begin{figure}[htbp]
\begin{center}
\includegraphics*[width=\columnwidth]{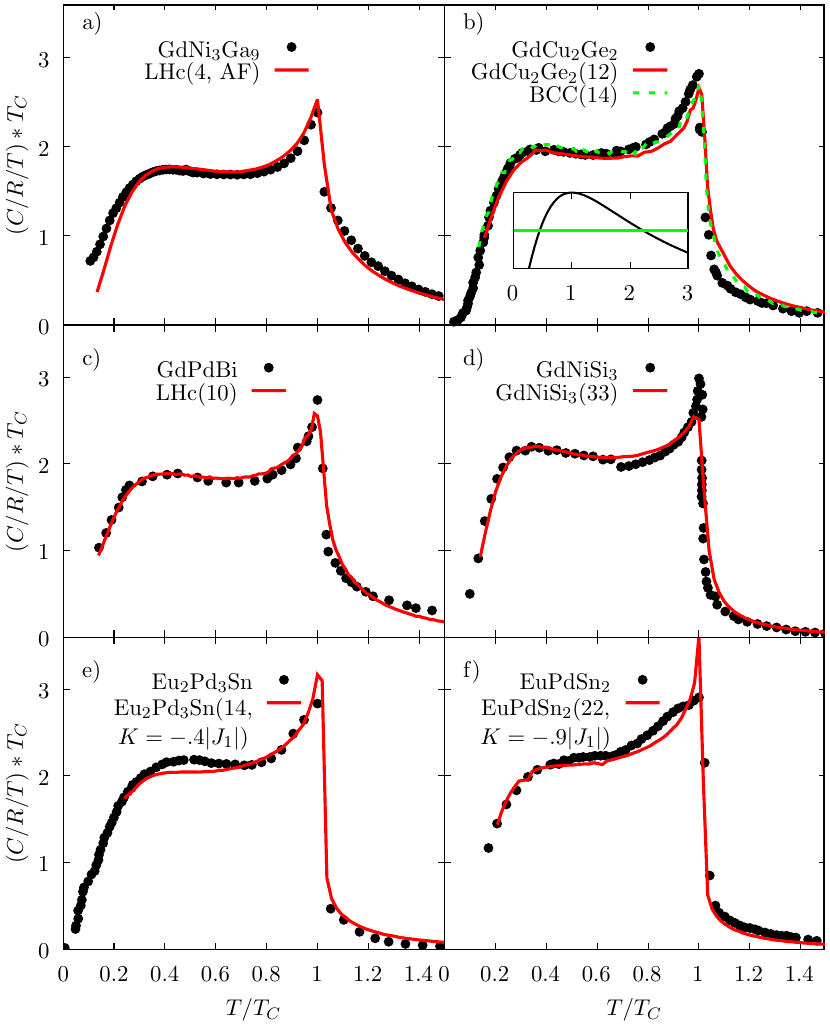}
\end{center}
\caption{ Experimental specific heat $C/T$ versus $T/T_N$ for six 4f$^7$ compounds (dots) and our theoretical fits (lines). $K = 0$ for all Gd-based compounds (GdNi$_3$Ga$_9$, GdCu$_2$Ge$_2$, GdPbBi, GdNiSi$_3$).
Inset for GdCu$_2$Ge$_2$  show $z$ as a function of $J_3/J_1$ (black line) and the fitting $z=14$ value (horizontal green line). 
Experimental data sources: Refs. \onlinecite{silva2017efeitos,nakamura2023discovery,Jesus14,duong2002correlation,Arantes18,Curlik18,Curlik24}. }
\label{exp}
\end{figure}

In this section, we present fits of the specific heat of six compounds: GdNi$_3$Ga$_9$, GdCu$_2$Ge$_2$, GdPbBi, GdNiSi$_3$, Eu$_2$Pd$_3$Sn$_3$, and EuPdSn$_2$ (see Figure \ref{exp}). 
The fitting procedure uses the experimentally determined critical temperature $T_N$ as a fixed input, with the effective number of neighbors $z$ and axial anisotropy $K$ as free parameters. For Gd compounds, $K=0$ was used as the anisotropy term was found to be negligible. The assumption of uniform exchange interactions ($J_\delta = J$) is justified by our theoretical results showing that different $J_\delta$ values with the same effective number of neighbors $z$ yield nearly identical specific heat curves (see Section \ref{compa}).

The quality of the fits is excellent. For comparison, a mean-field approach would fail to reproduce the specific heat curves, particularly the shape of the peak and the high-temperature tail (see mean-field specific heat in Figures \ref{cst} and \ref{cstmf}). For a quantitative measure
of the fits, see the discussion of Appendix \ref{app:distanceExperiment}. 
The $L^2$ distance between the experimental curve and the corresponding fit is in all cases $\ell_2 \leq 0.1$. 
Taking into account the experimental substraction of the phonon contribution, this is a reasonable distance.

The parameters of the fits for the different compounds are listed in Table \ref{tabpara}. 
The table also shows $z_0$.
It represents the minimum number of nearest neighbors required to maintain 3D connectivity in the reported crystal structure.
The effective number of neighbors $z$ accounts for both geometric coordination and relative interaction strengths, which can differ from the nominal $z_0$ due to longer-range interactions.

\begin{table}[h!]
    \centering
    \begin{tabular}{ l c c c c c l }
    \hline
    Compound & $T_N$ (exp,[K]) & $z_0$ & $z$(fit) & $\vert J\vert$[K] & $K$[K] \\ \hline
    GdNi$_3$Ga$_9$  & 20           & 4       & 4        & 1.5 & 0    \\ \hline
    GdPbBi                   & 13           & 12    & 10      & --$^\dagger$ & 0     \\ \hline 
    GdCu$_2$Ge$_2$  & 12           & 12    & 14      & 0.20$^\ddagger$& 0     \\ \hline
    GdNiSi$_3$            & 22           & 17    & 33      & 0.15 & 0        \\ \hline
    Eu$_2$Pd$_3$Sn  & 47            & 4      & 14     & 0.71 & -0.28 \\ \hline
    EuPdSn$_2$          & 12            & 8  & 22        & 0.11 & -0.09 \\ \hline
    \end{tabular}
    \caption{Parameters used in the fits: $ T_N $ is the N\'eel temperature, $ z_0 $ is the number of nearest neighbors in the reported structure (compatible with 3D connectivity), $ z $ (fit) is the effective number of neighbors, $ J $ is the interaction strength, and $ K $ is the anisotropy.
    $^\dagger$ From the fit we can only conclude that $|J_1|<|J_2|$, see main text.
    $^\ddagger$ This value correspond to $|J_1|=|J_2|=0.20K$ and $|J_3|=0.09K$, see main text.
    }
    \label{tabpara}
    \end{table}

GdNi$_3$Ga$_9$ \cite{silva2017efeitos,nakamura2023discovery} exhibits a layered honeycomb structure with $z_0 = 4$ (three intralayer and one interlayer nearest neighbors), consistent with $z = 4$ found in the fit. 
The specific heat fit (Figure \ref{exp}a) reproduces the experimental data with excellent accuracy. 
The low $z$ value indicates significant thermal and quantum fluctuations
accounting for the observed broadened specific heat peak.
As the LHc(4,AF) lattice (corresponding to an antiferromagnetic interaction) has $T_N/ J_1 \sim 13$ while experimentally $T_N=20K$ we can deduce $J \sim 20$K$/13\sim 1.5$K.
This result of a dominant first-neighbor antiferromagnetic interaction is  further supported by the observation of a small (1.8°) deviation from perfect AF alignment between magnetic moments in adjacent layers \cite{nakamura2023discovery}, which may arise from weak Dzyaloshinskii-Moriya interactions (DMI).

For the rest of the structures, ferromagnetic interactions are used for the fits.
As stated in Section \ref{nei}, as long as there is no significant frustration, the difference between ferromagnetic and antiferromagnetic interactions for $z>6$ is only noticeable at very low temperatures, well below the temperature range of interest in this study. 
In Fig. \ref{figure:ChangeJsGreaterJ1} we show the specific heat of a SC and a BCC lattice with AF interactions ($z=6,8$) that compares well with the FM case as $z$ increases.
The key features of the specific heat curves, such as the shape of the peak and the high-temperature tail, are primarily determined by the effective number of neighbors and the single-ion anisotropy. 
These features are largely insensitive to the sign of the exchange interaction in the absence of strong frustration.
For this reason we can also compare the experimental N\'eel temperature ($T_N$) with the theoretical Curie temperature ($T_C$).

GdCu$_2$Ge$_2$ crystallizes in a tetragonal structure \cite{duong2002correlation}, with the Gd sublattice adopting a body-centered tetragonal arrangement. 
Due to the strong elongation of the unit cell ($c/a \gg 1$), considering only the four first-nearest neighbors (NN) would result in a quasi-two-dimensional system, leading to a broadened magnetic transition near $T_N$. 
The four NNN lie in the same crystallographic plane as the NN, preserving the 2D character.
To recover the three-dimensional connectivity required to reproduce the experimental specific heat, at least the eight third-nearest neighbors must be included.
This implies that a minimum viable model must account for both the first and third neighbor interactions, yielding a total coordination number $z_0 = 12$. 
As discussed earlier, such a model resembles other lattices with $z=12$, such as FCC(12). 
In panel (b), we show that the specific heat is reasonably described by the GdCu$_2$Ge$_2$(12) lattice. 
However, the best agreement with experiment is achieved using the BCC(14) lattice (see panel b and Appendix \ref{app:distanceExperiment}). 
We therefore associate an effective coordination number $z = 14$ with this compound.
If, to simplify, we consider $J_1$ ($z_1=4$) equal to  $J_2$ ($z_2=4$) we get the equation
 \[
 1/z = \frac{(z_1+z_2) J_1^2 + z_3 J_3^2}{( (z_1+z_2) |J_1| + z_3 |J_3|)^2}.
 \] 
This assumption leads to the possibilities:
\begin{itemize}
 \item $|J_3/J_1|=2.21$ ($z=14$) which we can disregard has it seems unlikely  to have a larger strength associated to a more distant bond.
 \item $|J_3/J_1|=0.45$ ($z=14$) appears to be the only reasonable solution.
\end{itemize}
Without the actual simulation for this particular combination of couplings, we can only estimate the coupling strengths.
To do so, we use the mean-field result for the critical temperature (see section \ref{MF}).
\[
\frac{k_B T_C}{ \frac{S(S+1)}{3}} =   (z_1+z_2) |J_1| + z_3 |J_3| = (8  + 8 \times0.45  ) |J_1|
\]
from which we obtain $|J_1|=|J_2|\sim0.2$ K and $|J_3|\sim0.09$ K.
From these results we can not determine the signs of the coupling, but from the antiferromagnetic nature of the compound, one infers that at least one of these couplings should be antiferromagnetic.

GdPdBi adopts a half-Heusler structure with the Gd sublattice forming a face-centered cubic (FCC) lattice. Fitting with the layered honeycomb lattice (LHc(10), $z=10$) requires an unexpectedly large $J_2/J_1$ ratio of approximately $6.1$. 
To validate our approach, we compared our results with inelastic neutron scattering measurements on the related compound GdPtBi ($T_N = 8.5$ K)\cite{kreyssig2011magnetic,sukhanov2020magnon}. The analysis of GdPtBi\cite{sukhanov2020magnon} from spin-wave theory reported $J_1=0.14$ K, $J_2=0.27$ K, $J_3=0.049$ K, and $K=0.0098$ K, with $|J_2|/|J_1| > 1$ (specifically $|J_2|/|J_1| \approx 1.93$), demonstrating that second-neighbor interactions dominate in this compound and that the anisotropy is small ($|K|/|J_1|\lesssim 0.1$). In GdPtBi, the magnetic structure consists of ferromagnetic (111) planes antiferromagnetically stacked along [111]\cite{kreyssig2011magnetic,muller2014magnetic}, exhibiting low frustration.
This structural and magnetic context validated the plausibility of the $|J_2|/|J_1| \gtrsim1$ ratio for GdPdBi. 
But as shown in section \ref{compa} the specific heat for $|J_1|<|J_2|$ is almost independent of $J_2$ and does not allow us to determine a ratio of interactions.


The crystal structure of GdNiSi$_3$ is an orthorhombic system.
Gd atoms  forms a bilayered arrangement.
Only interaction at the 7th nearest neighbor connects bilayers to provide a tridimensional connectivity ($z_0=17$).
A good description of the experimental data is obtained considering up to 10th nearest-neighbor ($z=25$) while the best fit is seen considering up to 13th NN ($z=33$).
This suggests that long range interactions, well beyond first neighbors, are relevant in this compound.
Given the complexity of these interactions we only consider the uniform $J_\delta$ case. 
The GdNiSi$_3$(33) lattice has $T_C/{\vert J \vert} \sim 145$ while experimentally $T_N=22$K so we can deduce $\vert J \vert  \sim 22$K$/145=0.15$K. 
The large amount of couplings needed (up to 13 nearest-neighbors)  certainly come from a distribution of interactions.
This distribution could be determined with
first-principle calculations that considers at least this coordination.


This completes the Gd compounds that we have considered.
In a previous work \cite{Facio15}, the specific heat of   
GdCoIn$_5$ has been fitted with essentially the same procedure. The compound has a tetragonal structure. 
The authors have extrapolated to the thermodynamic limit the specific heat that corresponds to a simple cubic lattice with $z=6$ nearest neighbors with antiferromagnetic interactions, providing an excellent fit of the data.

Now turning to Eu compounds, the importance of $ K $ for these compounds may be attributed to the $ \sim 25\% $ larger ionic radius of Eu$^{2+} $ compared to Gd$^{3+} $ \cite{shannon1976revised}, which makes the former more susceptible to crystal-field effects.

The Eu sublattice of Eu$_2$Pd$_3$Sn can be thought of as a stacking of corrugated hexagonal layers in the ab plane.
The smaller coordination number to have a tridimensional connectivity is $z_0=4$. 
This involves up to the third coordination shell of nearest neighbors (first and
second nearest neighbors lie in the same layer).
To properly compare the specific heat of Eu$_2$Pd$_3$Sn \cite{Curlik24} with a Hamiltonian model  we have to consider up to 6th nearest-neighbor in this lattice leading to a coordination $z=14$.
The point group at the Eu sites is $C_{1h}$, which allows for the inclusion of the anisotropy term $H_A$, as discussed in section \ref{sec:Anisotropy}.
So an anisotropy term with  $K/\vert J_1\vert=-0.4$ was included.
With this anisotropy the Eu$_2$Pd$_3$Sn(14) lattice has $T_C/{\vert J_0 \vert} \sim 66$.
Experimentally $T_N=47$K so we can deduce, assuming uniform couplings, $\vert J \vert  \sim 47$K$/66=0.71$K.

Finally, the Eu sublattice of EuPdSn$_2$ can be thought of as a stacking of  rectangular  layers in the ac plane.
The smaller coordination number to have a tridimensional connectivity is $z_0=8$. This involves up to the third coordination shell of nearest-neighbor interactions.
To properly compare the specific heat of EuPdSn$_2$ \cite{Curlik18} with a Hamiltonian model  we have to consider up to 7th nearest-neighbor in this lattice leading to a coordination $z=22$.
The point group at the Eu sites is $C_{2v}$, which also allows for the inclusion of the term $H_A$.
So an anisotropy term with  $K/\vert J_0\vert=-0.9$ was included.
With this anisotropy the EuPdSn$_2$ lattice has $T_C/{\vert J_0 \vert} \sim 112$.
Experimentally $T_N=12$K from which we deduce $\vert J \vert  \sim 12$K$/112=0.11$K. 
The discrepancies near  $T_N$  likely originate from factors not explicitly included in our model, such as more complex magnetic ordering (e.g., a collinear amplitude-modulated structure) resulting from modulated interactions and crystal-field effects, as discussed in Ref. \onlinecite{BlancoII}. 

Since all the mentioned compounds contain transition-metal ions, which may be magnetic, one might question their influence on the specific heat. However, our analysis shows that the entropy obtained by integrating $ C/T $ is close to $ \ln(8) $ per rare-earth ion (consistent with $ S = 7/2 $), indicating that any effects from transition metals are negligible.

While several theoretical curves may appear visually similar to the experimental data, only a single combination of  $z$ and  $K$  simultaneously minimizes the  $L^2$  distance and faithfully reproduces the low-temperature specific heat, enabling unambiguous extraction of microscopic parameters.

The success of these fits suggests that the effective number of neighbors $z$ is the dominant parameter determining the specific heat of 4f$^7$ systems, with axial anisotropy $K$ providing additional refinement for Eu compounds. This approach significantly outperforms mean-field theory, which would fail to reproduce the observed specific heat curves (see Fig. \ref{cst}).

\section{Summary}
\label{sum}

We present a unified framework for understanding the specific heat of 4f$^7$ magnetic compounds (Gd$^{3+}$ and Eu$^{2+}$), characterized by a ground state multiplet $^{8}S_{7/2}$.
Through extensive quantum Monte Carlo simulations, we demonstrate that the specific heat is primarily governed by two parameters: the effective number of neighbors $z$ and axial anisotropy $K$.

The effective  number of neighbors $z$ quantifies the strength of thermal and quantum fluctuations, with larger $z$ values corresponding to reduced fluctuations and specific heat curves closer to mean-field predictions. Crucially, we find that the specific heat is remarkably insensitive to the detailed lattice structure for a given $z$ as long as the couplings do not strongly increase with distance. 
This enables a universal description of diverse magnetic systems. 

The origin of the anisotropy $K$ lies in a small admixture of $^{6}P_{7/2}$ in the ground-state multiplet. This admixture introduces a component with $L = 1$, rendering the system sensitive to axial crystal fields.

We successfully fit the specific heat of six experimental compounds (four Gd-based and two Eu-based) using only $z$ and $K$ as free parameters, with excellent agreement between theory and experiment. Another Gd compound has been studied before \cite{Facio15}.
The present approach allowed us to extract physical information not accessible through mean field.
Notably, axial anisotropy is essential for Eu compounds but can be neglected for Gd compounds, reflecting the larger ionic radius of Eu$^{2+}$ and its greater susceptibility to crystal-field effects.

This work provides a simple yet powerful tool for interpreting specific heat data in 4f$^7$ systems, significantly outperforming mean-field approaches and enabling the extraction of (effective) fundamental magnetic interaction parameters from experimental measurements.
While our two-parameter model works well for most of the compounds studied, it may require additional parameters for systems with strong frustration, complex magnetic order, or with interactions strengths that increases with distance as in the case of GdPtBi. 

Experimental researchers studying new 4f$^7$ compounds can now follow this straightforward procedure: (1) Measure the specific heat and determine $T^*$, the critical N\'eel or Curie temperature; (2) Compare the $(C/T)*T^*$ vs $T/T^*$ curve to our reference data in Ref.  \cite{note}; (3) Identify the $z$ value that provides the best match; (4) Use Eq. (\ref{Rdefinicion}) and known crystal structure to constrain possible exchange interaction patterns.
This approach provides physically meaningful constraints on magnetic interactions.
These results could also guide the development of more accurate theoretical models for compounds.

\begin{acknowledgments}
The authors are grateful to I. \v{C}url\'{\i}k, L. de Sousa Silva, J.G.S. Duque, P.G. Pagliuso, and M. Avila  for allowing us access to the original experimental  data.
D.J.G. is supported by PIP 11220200101796CO of CONICET, Argentina.
Computational resources were provided by the HPC cluster of the Physics Department at Centro Atómico Bariloche (CNEA) and Clementina XXI.
Language editing was supported by the use of Qwen3\cite{qwen3technicalreport}, with final approval of all content by the authors.
\end{acknowledgments}

\appendix

\section{Mean value and fluctuations of a random variable}
\label{fluc}

Here we calculate the mean value and fluctuations of a random variable $x$ with different intensities \footnote{The same result can be obtained considering each bond as an independent variable.}. 
For simplicity we consider only two intensities 
$J_{1}$ and $J_{2}$. Extension to the general case is straightforward. We
consider $z_{1}$ attempts in which the probability of the result $J_{1}$ is 
$p$ and that of $-J_{1}$ is $q=1-p$. In addition, there are $z_{2}$ attempts
with probability $r$ of finding the result $J_{2}$ and $s=1-r$ for the
result $-J_{2}$. At the end we will take $r=p$.
For non-frustrated systems, both $J_\delta$ can be chosen positive. 
Clearly, the expectation
value of $x$ is

\begin{equation}
\langle x\rangle =z_{1}J_{1}(p-q)+z_{2}J_{2}(r-s).  \label{mv}
\end{equation}

On the other hand, considering the probability of each individual event, we have:

\begin{eqnarray}
\langle x^{2}\rangle &=&\sum\limits_{n=0}^{z_{1}}\sum\limits_{m=0}^{z_{2}}
\binom{z_{1}}{n}\binom{z_{2}}{m}p^{n}q^{z_{1}-n}r^{m}s^{z_{2}-m} \notag \\
&&\times \left[nJ_{1}-(z_{1}-n)J_{1}+mJ_{2}-(z_{2}-m)J_{2}\right]^2.
\label{mv2}
\end{eqnarray}

Using the operator

\begin{eqnarray}
&&O=\left[ J_{1}\left( p\frac{\partial }{\partial p}-q\frac{\partial }
{\partial q}\right) +J_{2}\left( r\frac{\partial }{\partial r}-
s\frac{\partial }{\partial s}\right) \right] ,\notag \\
&&\mathrm{ and }(p+q)^{z_{1}}=\sum\limits_{n=0}^{z_{1}}\binom{z_{1}}{n}p^{n}q^{z_{1}-n},  \label{op}
\end{eqnarray}
and a similar expression for $(r+s)^{z_{2}}$, 
where $p$ and $q$ as well as $r$ and $s$ are taken as independent variables in the derivatives, 
one realizes that the
expectation value Eq. (\ref{mv2}) can be written as

\begin{equation}
\langle x^{2}\rangle =O^{2}(p+q)^{z_{1}}(r+s)^{z_{2}}.  \label{mvo}
\end{equation}
Performing the calculation, we find after some algebra

\begin{eqnarray}
\langle x^{2}\rangle  &=&J_{1}^{2}z_{1}\left[ 1+(z_{1}-1)(p-q)^{2}\right] \notag \\
&&+J_{2}^{2}z_{2}\left[ 1+(z_{2}-1)(r-s)^{2}\right]   \notag \\
&&+2J_{1}J_{2}z_{1}z_{2}(p-q)(r-s)  \label{mvx2}
\end{eqnarray}

Using Eqs. (\ref{mv}) and (\ref{mvx2}), setting $r=p$ and replacing
$q=s=1-p$, one finally obtains

\begin{equation}
\frac{\langle x^{2}\rangle -\langle x\rangle ^{2}}{\langle x\rangle ^{2}}
=\frac{4p(1-p)}{(2p-1)^2}\frac{z_{1}J_{1}^{2}+z_{2}J_{2}^{2}}{\left(
z_{1}J_{1}+z_{2}J_{2}\right) ^{2}}.  \label{ratio}
\end{equation}

Naturally, if $\langle x \rangle=0 $ as it is the case for temperatures above the critical temperature $T_C$,
Eq. (\ref{ratio}) gives  a divergent result.
In this case a meaningful comparison should involve the fluctuations above $T_C$ and the value of $\langle x \rangle=0 $ at some point below
$T_C$.

\section{Distance between specific heat curves} \label{app:distance}

To quantify the similarity between specific heat curves across different lattices, we define a comparison interval that excludes regions dominated by spin-wave excitations and the critical region. Specifically, we discard temperatures below $T/T_C = 0.25$, where low-energy magnons dominate, and within the critical region $0.95 < T/T_C < 1.05$, where finite-size and scaling effects obscure universal behavior. Above the transition, we restrict our analysis to $T \leq 1.25\,T_C$, as above this temperature the specific heat decays rapidly and contributes negligibly to the integrated signal.

The selected interval $[0.25, 1.25]$ is divided into 50 equally spaced bins. For each curve, the mean value of the specific heat points falling within each bin is computed. If a curve has no data points in a given bin, that bin is excluded from the comparison for that pair of curves.

The distance between two specific heat curves, $C_1(T)$ and $C_2(T)$, is then defined as the normalized $L^2$ distance over the set of valid bins:
\[
\ell_2 = \sqrt{ \frac{1}{n_{\text{valid}}} \sum_{i=1}^{n_{\text{valid}}} \left( \langle C_1 \rangle_i - \langle C_2 \rangle_i \right)^2 },
\]
where $\langle C_k \rangle_i$ denotes the mean value of curve $k$ in bin $i$, and $n_{\text{valid}}$ is the number of bins for which both curves have at least one data point. This metric provides a robust measure of dissimilarity between specific heat profiles across lattices, insensitive to small sampling gaps.

\section{Similarity of specific heat curves for uniform exchange couplings} \label{app:distanceJsUniform}

\begin{figure*}[htbp]  
\begin{center}  
\includegraphics*[width=\textwidth]{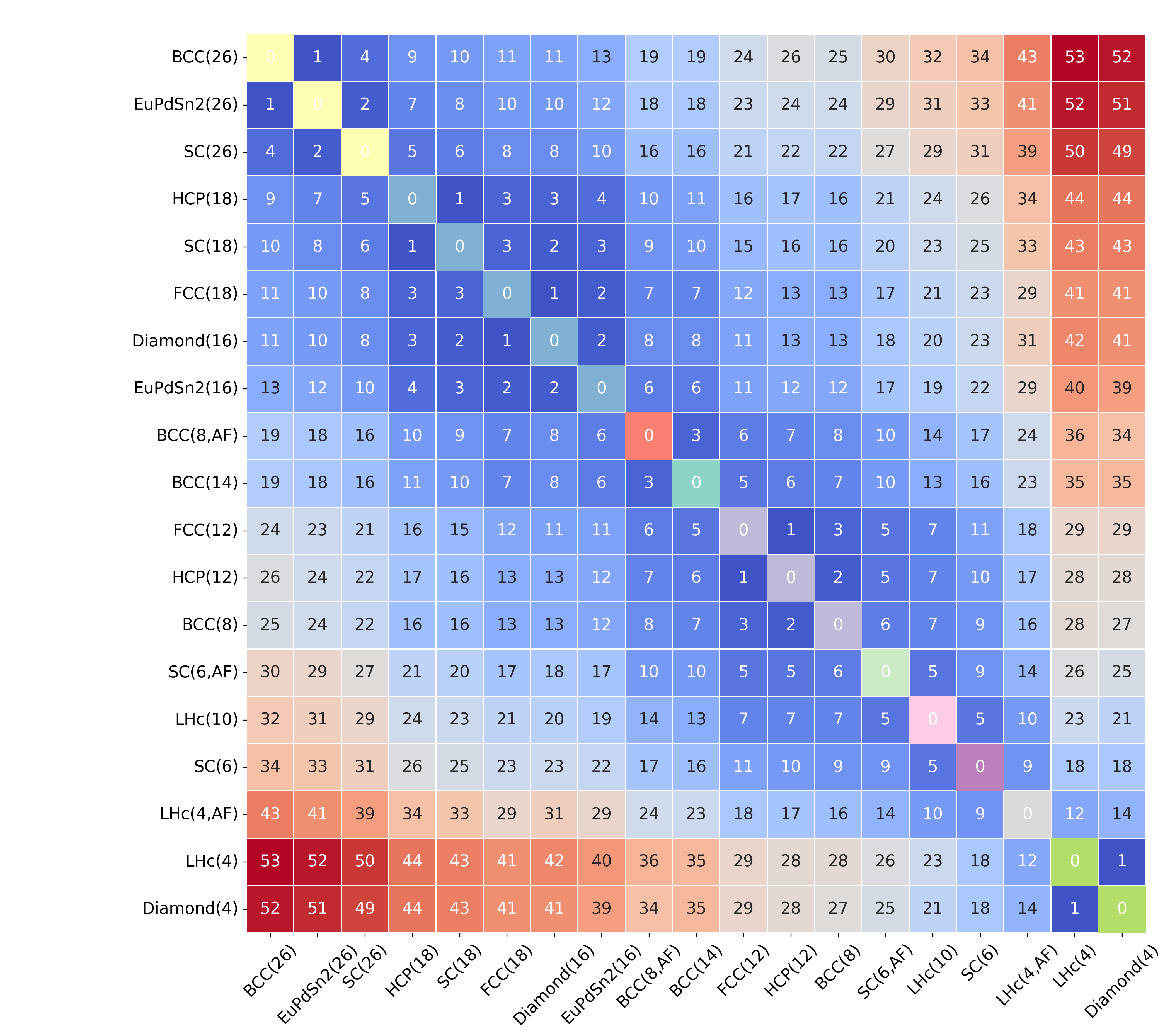}  
\end{center}  
\caption{  
$L^2$ distance matrix between specific heat curves $C(T)/T$ for 19 non-frustrated lattices with uniform exchange couplings ($J_\delta = J_1$). Off-diagonal elements show the dissimilarity $100 \times \ell_2$ (i.e., multiplied by 100 for visual clarity), computed over the range $0.25 \leq T/T_C \leq 1.25$ (see Appendix~\ref{app:distance}). Colors range from blue (low distance, high similarity) to red (high distance, low similarity).
Curves with $\ell_2 < 0.03$ are grouped into the same cluster.
}  
\label{figure:HeatMapJsIguales}  
\end{figure*}

Figure~\ref{figure:HeatMapJsIguales} presents the $L^2$ distance matrix between specific heat curves $C(T)/T$ for 19 non-frustrated lattices with uniform exchange couplings (i.e., $J_\delta = J_1$), as listed in Table~\ref{Table:CriticalTemperatures}. The matrix combines a numerical table with a color heatmap, where off-diagonal entries represent the distance $\ell_2$ between pairs of curves, computed over the temperature range $0.25 \leq T/T_C \leq 1.25$ (see Appendix~\ref{app:distance}). 

The color scale ranges from blue (lowest distance, $\ell_2 \approx 0$) to red (highest distance), with the diagonal entries indicating the clustering structure of lattices by their effective coordination number $z$. The rows and columns are ordered such that lattices with similar $z$ values are grouped together, resulting in a clear block-diagonal structure. This ordering closely follows the sequence of increasing $z$, as defined in Eq.~(\ref{Rdefinicion}).

Lattices with $\ell_2 < 0.03$ are considered to belong to the same cluster, as this threshold corresponds approximately to the visual discriminability limit of curves in Figure~\ref{cst} and to the estimated numerical uncertainty in our QMC data. 

For example, a distinct yellow cluster emerges at $z = 26$, comprising the BCC, EuPdSn$_2$, and simple cubic (SC) lattices. Similarly, lattices with $z = 18$ (HCP, SC, FCC) and $z = 16$ (diamond, EuPdSn$_2$) form well-defined clusters. We also observe clustering between $z = 8$ (BCC) and $z = 12$ (FCC, HCP), as well as between $z = 4$ (diamond, LHc), further supporting the universality of $z$ as the primary determinant of specific heat behavior under uniform couplings.

\section{Similarity of specific heat curves for non-uniform exchange couplings}  \label{app:distanceJsNonUniform}

For systems with non-uniform exchange couplings ($J_2 \neq J_1$), the $L^2$ distance $\ell_2$ between specific heat curves quantifies deviations from the universality of $z$. Tables~\ref{tab:J2_less_J1} and \ref{tab:J2_greater_J1} summarize pairwise distances for lattices with $J_2 < J_1$ and $J_2 > J_1$, respectively, computed over the range $0.25 \leq T/T_C \leq 1.25$. 

When $|J_2| < |J_1|$ (Table~\ref{tab:J2_less_J1}), curves with the same $z$ remain highly similar ($\ell_2 < 0.06$), consistent with the dominance of $z$ as the controlling parameter. The most notable exception is the layered honeycomb lattice LHc(10), which exhibits slightly larger distances ($\ell_2 \approx 0.07$–$0.11$) relative to cubic lattices, reflecting its distinct low-dimensional topology.

When $|J_2| \gg |J_1|$ (Table~\ref{tab:J2_greater_J1}), deviations increase significantly for small $z$ ($z=8$, $\ell_2 \sim 0.1$–$0.23$), but diminish as $z$ increases. For $z \geq 14$, distances fall below 0.07, indicating that the universality of $z$ is restored when the effective number of neighbors becomes sufficiently large, even under strong hierarchical couplings.

\begin{table}[htbp]
\centering
\caption{$L^2$ distances ($\ell_2$) between specific heat curves for lattices with $|J_2| < |J_1|$.}
\label{tab:J2_less_J1}
\begin{tabular}{lll}
\hline
$z=6$ & \multicolumn{2}{c}{$\ell_2 = 0.05$} \\
\hline
\multicolumn{2}{l}{SC(6) \quad vs \quad Diamond($J_2 = -0.080$)} & 0.05 \\
\hline
$z=8$ & \multicolumn{2}{c}{$\ell_2 \leq 0.06$} \\
\hline
\multicolumn{2}{l}{BCC(8) \quad vs \quad SC($J_2 = -0.082$)} & 0.03 \\
\multicolumn{2}{l}{Diamond($J_2 = -0.155$) \quad vs \quad SC($J_2 = -0.082$} & 0.04 \\
\multicolumn{2}{l}{BCC(8) \quad vs \quad Diamond($J_2 = -0.155$)} & 0.06 \\
\hline
$z=10$ & \multicolumn{2}{c}{$\ell_2 \leq 0.11$} \\
\hline
\multicolumn{2}{l}{Diamond($J_2 = -0.236$) \quad vs \quad SC($J_2 = -0.16$)} & 0.03 \\
\multicolumn{2}{l}{Diamond($J_2 = -0.236$) \quad vs \quad BCC($J_2 = -0.17$)} & 0.04 \\
\multicolumn{2}{l}{SC($J_2 = -0.16$) \quad vs \quad BCC($J_2 = -0.17$)} & 0.05 \\
\multicolumn{2}{l}{LHc(10) \quad vs \quad SC($J_2 = -0.16$)} & 0.07 \\
\multicolumn{2}{l}{LHc(10) \quad vs \quad Diamond($J_2 = -0.236$)} & 0.08 \\
\multicolumn{2}{l}{LHc(10) \quad vs \quad BCC($J_2 = -0.17$)} & 0.11 \\
\hline
$z=12$ & \multicolumn{2}{c}{$\ell_2 \leq 0.04$} \\
\hline
\multicolumn{2}{l}{SC($J_2 = -0.25$) \quad vs \quad BCC($J_2 = -0.40$)} & 0.02 \\
\multicolumn{2}{l}{Diamond($J_2 = -0.333$) \quad vs \quad BCC($J_2 = -0.40$)} & 0.02 \\
\multicolumn{2}{l}{FCC(12) \quad vs \quad Diamond($J_2 = -0.333$)} & 0.02 \\
\multicolumn{2}{l}{Diamond($J_2 = -0.333$, $z=12$) \quad vs \quad SC($J_2 = -0.25$)} & 0.03 \\
\multicolumn{2}{l}{FCC(12) \quad vs \quad BCC($J_2 = -0.40$)} & 0.03 \\
\multicolumn{2}{l}{FCC(12) \quad vs \quad SC($J_2 = -0.25$)} & 0.04 \\
\hline
$z=14$ & \multicolumn{2}{c}{$\ell_2 \leq 0.06$} \\
\hline
\multicolumn{2}{l}{BCC(14) \quad vs \quad FCC($J_2 = -0.18$)} & 0.02 \\
\multicolumn{2}{l}{SC($J_2 = -0.35$) \quad vs \quad FCC($J_2 = -0.18$)} & 0.02 \\
\multicolumn{2}{l}{BCC(14) \quad vs \quad SC($J_2 = -0.35$)} & 0.03 \\
\multicolumn{2}{l}{BCC(14) \quad vs \quad Diamond($J_2 = -0.472$)} & 0.04 \\
\multicolumn{2}{l}{Diamond($J_2 = -0.472$) \quad vs \quad FCC($J_2 = -0.18$)} & 0.05 \\
\multicolumn{2}{l}{Diamond($J_2 = -0.472$) \quad vs \quad SC($J_2 = -0.35$)} & 0.06 \\
\hline
$z=16$ & \multicolumn{2}{c}{$\ell_2 \leq 0.04$} \\
\hline
\multicolumn{2}{l}{SC($J_2 = -0.50$) \quad vs \quad FCC($J_2 = -0.40$)} & 0.02 \\
\multicolumn{2}{l}{Diamond(16) \quad vs \quad FCC($J_2 = -0.40$)} & 0.04 \\
\multicolumn{2}{l}{Diamond(16) \quad vs \quad SC($J_2 = -0.50$)} & 0.04 \\
\hline
$z=18$ & \multicolumn{2}{c}{$\ell_2 \leq 0.04$} \\
\hline
\multicolumn{2}{l}{FCC(18) \quad vs \quad BCC($J_1 = J_2, J_3 = -0.175$)} & 0.02 \\
\multicolumn{2}{l}{SC(18) \quad vs \quad BCC($J_1 = J_2, J_3 = -0.175$)} & 0.04 \\
\hline
\end{tabular}
\end{table}

\begin{table}[htbp]
\centering
\caption{$L^2$ distances ($\ell_2$) between specific heat curves for lattices with $|J_2| \gg |J_1|$.}
\label{tab:J2_greater_J1}
\begin{tabular}{lll}
\hline
$z=8$ & \multicolumn{2}{c}{$\ell_2 = 0.10$–$0.23$} \\
\hline
\multicolumn{2}{l}{BCC($J_2 = -8$) \quad vs \quad FCC($J_2 = -12.32$)} & 0.10 \\
\multicolumn{2}{l}{Diamond($J_2 = -0.155$) \quad vs \quad BCC($J_2 = -8$)} & 0.13 \\
\multicolumn{2}{l}{Diamond($J_2 = -0.155$) \quad vs \quad FCC($J_2 = -12.32$)} & 0.23 \\
\hline
$z=10$ & \multicolumn{2}{c}{$\ell_2 = 0.09$–$0.18$} \\
\hline
\multicolumn{2}{l}{Diamond($J_2 = -0.236$) \quad vs \quad BCC($J_2 = -3.82$)} & 0.09 \\
\multicolumn{2}{l}{BCC($J_2 = -3.82$) \quad vs \quad FCC($J_2 = -6.1$)} & 0.10 \\
\multicolumn{2}{l}{Diamond($J_2 = -0.236$) \quad vs \quad FCC($J_2 = -6.1$)} & 0.18 \\
\hline
$z=12$ & \multicolumn{2}{c}{$\ell_2 = 0.08$–$0.17$} \\
\hline
\multicolumn{2}{l}{Diamond($J_2 = -0.333$) \quad vs \quad BCC($J_2 = -2.27$)} & 0.08 \\
\multicolumn{2}{l}{BCC($J_2 = -2.27$) \quad vs \quad FCC($J_2 = -4$)} & 0.11 \\
\multicolumn{2}{l}{Diamond($J_2 = -0.333$) \quad vs \quad FCC($J_2 = -4$)} & 0.17 \\
\hline
$z=14$ & \multicolumn{2}{c}{$\ell_2 < 0.13$} \\
\hline
\multicolumn{2}{l}{Diamond($J_2 = -3.527$) \quad vs \quad SC($J_2 = -5.64$)} & 0.03 \\
\multicolumn{2}{l}{Diamond($J_2 = -3.527$) \quad vs \quad Diamond($J_2 = -0.472$)} & 0.04 \\
\multicolumn{2}{l}{Diamond($J_2 = -0.472$) \quad vs \quad SC($J_2 = -5.64$)} & 0.06 \\
\multicolumn{2}{l}{FCC($J_2 = -2.82$) \quad vs \quad SC($J_2 = -5.64$)} & 0.07 \\
\multicolumn{2}{l}{Diamond($J_2 = -3.527$) \quad vs \quad FCC($J_2 = -2.82$)} & 0.08 \\
\multicolumn{2}{l}{FCC($J_2 = -2.82$) \quad vs \quad Diamond($J_2 = -0.472$)} & 0.13 \\
\hline
$z=16$ & \multicolumn{2}{c}{$\ell_2 < 0.08$} \\
\hline
\multicolumn{2}{l}{Diamond(16) \quad vs \quad SC($J_2 = -2.5$)} & 0.04 \\
\multicolumn{2}{l}{SC($J_2 = -2.5$) \quad vs \quad FCC($J_2 = -2$)} & 0.04 \\
\multicolumn{2}{l}{Diamond(16) \quad vs \quad FCC($J_2 = -2$)} & 0.07 \\
\hline
\end{tabular}
\end{table}

\section{Distances between experimental and theoretical specific heat curves}  \label{app:distanceExperiment}

To quantify the quality of the fits presented in Section~\ref{expe}, we compute the $L^2$ distance $\ell_2$ between each experimental specific heat curve and its theoretical counterparts, using the same temperature range $0.25 \leq T/T_N \leq 1.25$ (see Appendix~\ref{app:distance}). Table~\ref{tab:ExpTheoDistances} summarizes these distances for all candidate lattices with $\ell_2 < 0.3$, along with relevant anisotropy values where applicable. While the smallest $\ell_2$ values identify the closest matches, our selection of the optimal model for each compound prioritizes physical plausibility, especially the reproduction of the low-temperature behavior ($T < 0.6\,T_N$), which is least affected by uncertainties in phonon subtraction and experimental background correction.

\begin{table}[htbp]
\centering
\caption{$L^2$ distances ($\ell_2$) between experimental specific heat curves and theoretical predictions for 4f$^7$ compounds. Only models with $\ell_2 < 0.3$ and physically relevant coordination numbers or anisotropy values are shown. For Gd-based compounds, over 20 lattice models were evaluated; for Eu-based compounds, more than 30 combinations of $z$ and $K$ were considered. The selected model for each compound is indicated in bold.}
\label{tab:ExpTheoDistances}
\begin{tabular}{lll}
\hline
Compound & Theoretical lattice & $\ell_2$ \\
\hline
GdNi$_3$Ga$_9$ & {\bf LHc(4,AF)} & {\bf 0.07} \\
               & LHc(4)    & 0.12 \\
               & LHc(10)   & 0.15 \\
\hline
GdCu$_2$Ge$_2$ & FCC(18)      & 0.07 \\
               & {\bf BCC(14) }      & {\bf 0.09 } \\
               & GdCu$_2$Ge$_2$(16) & 0.11 \\
               & GdCu$_2$Ge$_2$(12) &  0.17 \\
\hline
GdPdBi         & SC(6)        & 0.07 \\
               & BCC(8)       & 0.07 \\
               & {\bf LHc(10)}  & {\bf 0.09} \\
               & FCC(12)      & 0.09 \\
               & FCC(18)      & 0.19 \\
\hline
GdNiSi$_3$     & {\bf GdNiSi$_3$(33)} & {\bf 0.06} \\
               & GdNiSi$_3$(25) & 0.17 \\
               & BCC(26)      & 0.06 \\
               & SC(18)       & 0.10 \\
               & BCC(14)      & 0.15 \\
\hline
Eu$_2$Pd$_3$Sn & Eu$_2$Pd$_3$Sn(14, $K = -0.2|J_1|$) & 0.09 \\
               & {\bf Eu$_2$Pd$_3$Sn(14, $K = -0.4|J_1|$)} & {\bf 0.09} \\
               & Eu$_2$Pd$_3$Sn(14, $K = -0.6|J_1|$) & 0.10 \\
               & Eu$_2$Pd$_3$Sn(14, $K = -0.8|J_1|$) & 0.18 \\
               & Eu$_2$Pd$_3$Sn(14, $K = 0$) & 0.20 \\
               & BCC(26)      & 0.12 \\
               & FCC(18)      & 0.15 \\
               & Diamond(16)  & 0.16 \\
               & Eu$_2$Pd$_3$Sn(20, $K = 0$) & 0.16 \\
\hline
EuPdSn$_2$     & EuPdSn$_2$(22, $K = -1.5|J_1|$) & 0.08 \\
               & {\bf EuPdSn$_2$(22, $K = -0.9|J_1|$)} & {\bf 0.10} \\
               & EuPdSn$_2$(22, $K = -1.2|J_1|$) & 0.10 \\
               & EuPdSn$_2$(22, $K = -0.6|J_1|$) & 0.14 \\
               & EuPdSn$_2$(22, $K = 0$) & 0.29 \\
               & EuPdSn$_2$(16, $K = -0.9|J_1|$) & 0.10 \\
               & EuPdSn$_2$(16, $K = -0.6|J_1|$) & 0.12 \\
               & EuPdSn$_2$(16, $K = 0$) & 0.30 \\
               & BCC(26)      & 0.22 \\
               & Diamond(16)  & 0.30 \\
\hline
\end{tabular}
\end{table}

\begin{itemize}
    \item \textbf{GdNi$_3$Ga$_9$}: Best described by the antiferromagnetic layered honeycomb lattice \textbf{LHc(4,AF)}, consistent with its magnetic structure described in Section \ref{nei} and dominant nearest-neighbor interaction $J_1$. The small $\ell_2$ confirms that long-range interactions are negligible.

    \item \textbf{GdCu$_2$Ge$_2$}: Although FCC(18) yields the smallest $\ell_2 = 0.07$, it fails to reproduce the low-temperature specific heat shape. We select \textbf{BCC(14)} ($\ell_2 = 0.09$), which accurately captures the experimental curve in the regime least affected by phonon subtraction artifacts.

    \item \textbf{GdPbBi}: The layered honeycomb lattice \textbf{LHc(10)} provides the most consistent description across $T/T_N < 0.6$, despite SC(6) and BCC(8) having similar $\ell_2$. The systematic deviation of SC(6) and BCC(8) below $T_N$ renders them less physically plausible for this compound.

    \item \textbf{GdNiSi$_3$}: Both \textbf{GdNiSi$_3$(33)} ($\ell_2 = 0.06$) and BCC(26) ($\ell_2 = 0.06$) reproduce the data excellently. The sharp feature near $T/T_N \approx 0.8$ is likely an experimental artifact, as no theoretical model reproduces it. We select GdNiSi$_3$(33) because it matches the compound crystal structure.

    \item \textbf{Eu$_2$Pd$_3$Sn}: Models with $K=0$ suggest an effective coordination number $z \approx 16$–$26$, but only those incorporating axial anisotropy reproduce the suppression of low-temperature entropy. Among the $z=14$ variants, values $K = -0.2|J_1|$ to $K = -0.6|J_1|$ yield $\ell_2 < 0.10$. We adopt $K = -0.4|J_1|$ as the representative value.

    \item \textbf{EuPdSn$_2$}: Models with $K=0$ are inconsistent with the experimental data across all coordination numbers. The best fits are obtained for the $z=22$ lattice with axial anisotropy $K = -0.9|J_1|$ ($\ell_2 = 0.10$), which accurately reproduces both the low-temperature entropy and the shape of the specific heat peak. Although the $z=16$, $K = -0.9|J_1|$ and $z=22$, $K = -1.5|J_1|$ models yield equal or slightly lower $\ell_2$ values of $0.10$ and $0.08$, respectively, they both underestimate the low-temperature specific heat. The complete failure of all $K=0$ models confirms that axial anisotropy is essential for describing the magnetic thermodynamics of Eu$^{2+}$ compounds.

\end{itemize}

These selections underscore that while minimal $\ell_2$ provides a useful quantitative guide, the physical consistency to the data remains paramount in identifying the correct magnetic model.

\bibliography{ref.bib}

\end{document}